\documentclass[useAMS,usenatbib,usegraphicx]{mn2e}
\usepackage{aas_macros}
\usepackage{amsmath}
\usepackage{amssymb}
\usepackage{booktabs,caption,fixltx2e}
\usepackage{url}
\usepackage{upgreek}
\usepackage{appendix}

\title[The Phase Modulation Method]{Finding binaries from phase modulation of pulsating stars with Kepler. IV. Detection limits and radial velocity verification}

\author[Simon J. Murphy et al.] 
{Simon J. Murphy$^{1,2,3,\dagger}$, Hiromoto Shibahashi$^3$ and Timothy R. Bedding$^{1,2}$\\
$^1$Sydney Institute for Astronomy (SIfA), School of Physics, The University of Sydney, NSW 2006, Australia\\
$^2$Stellar Astrophysics Centre, Department of Physics and Astronomy, Aarhus University, DK-8000 Aarhus C, Denmark\\
$^{3}$Department of Astronomy, The University of Tokyo, Tokyo 113-0033, Japan\\
\\
$^{\dagger}$email: simon.murphy@sydney.edu.au
}
 
\begin{document}

\maketitle

\begin{abstract}
We explore the detection limits of the phase modulation (PM) method of finding binary systems among multi-periodic pulsating stars. The method is an attractive way of finding non-transiting planets in the habitable zones of intermediate mass stars, whose rapid rotation inhibits detections via the radial velocity (RV) method. While oscillation amplitudes of a few mmag are required to find planets, many $\delta$\,Scuti stars have these amplitudes. In sub-optimal cases where the signal-to-noise of the oscillations is lower, low-mass brown dwarfs ($\sim$13\,M$_{\rm Jup}$) are detectable at orbital periods longer than about 1\,yr, and the lowest mass main-sequence stars (0.1--0.2\,M$_{\odot}$) are detectable at all orbital periods where the PM method can be applied. We use purpose-written Markov chain Monte Carlo (MCMC) software for the calculation of the PM orbits, which offers robust uncertainties for comparison with RV solutions. Using \textit{Kepler} data and ground-based RVs, we verify that these two methods are in agreement, even at short orbital periods where the PM method undersamples the orbit. We develop new theory to account for the undersampling of the time delays, which is also necessary for the inclusion of RVs as observational data in the MCMC software. We show that combining RVs with time delays substantially refines the orbits because of the complementarity of working in both the spatial (PM) and velocity (RV) domains simultaneously. Software outputs were tested through an extensive hare and hounds exercise, covering a wide range of orbital configurations including binaries containing two pulsators.
\end{abstract}



\section{Introduction}

Two methods are traditionally used to detect and characterise binary companions to stars: radial velocities (RVs) from spectroscopy and eclipse measurements from photometry. The availability of four years of high-precision photometry from the \textit{Kepler} Mission has facilitated a third method, namely measuring the effect of binary motion on stellar pulsations. This can be done via the frequency modulation (FM) method \citep{shibahashi&kurtz2012,shibahashietal2015,kurtzetal2015a}, or the phase modulation (PM) method \citep{murphyetal2014,koen2014,balona2014b}. Here we discuss the latter, which uses periodic phase shifts in stellar pulsations to infer a binary companion. Orbits are characterised using the time delays (sometimes called R\o{}mer delays) in place of RVs \citep{teltingetal2012}. Time delays have already been used to detect planetary companions to pulsating subdwarf B stars \citep{silvottietal2007}, building upon the traditional $O-C$ techniques that have been particularly successful with pulsars \citep{wolszczan&frail1992}. We have found the PM method to work well for $\delta$\,Scuti pulsating stars using \textit{Kepler} photometry.

All the aforementioned methods have advantages and disadvantages and are therefore complementary. RVs can reveal companions down to planetary masses in stars that have sharp spectral lines, but require a lot of observing time. Eclipses and transits push down to small planetary companions, provided the inclination is favourable, but geometry limits the number of systems observed to eclipse to about 1\:per\:cent. The PM method works best at longer orbital periods where the time delays are larger, but is restricted to stars with stable pulsations.

This is the fourth in a series of papers dedicated to development of the PM method. The first \citep{murphyetal2014} described the principle of obtaining the time delays from observed phase shifts of the stellar pulsations. The second \citep{murphy&shibahashi2015} provided an analytical method for fully solving the orbit, even in highly eccentric cases. The methodology contained within those papers is summarised in Sect.\,\ref{ssec:hound}, along with a description of new Markov chain Monte Carlo (MCMC) software used to determine the orbital parameters. A third study was recently made by \citet{comptonetal2016} to determine which kind of oscillating stars are suitable for PM analyses. They found that $\delta$\,Sct stars and white dwarfs were most favourable (cf. \citealt{dalessioetal2015}). However, we note that only around 20 white dwarfs were observed by Kepler during the main mission, and 14 of those were non-pulsators \citep{maozetal2015}, while thousands of $\delta$\,Sct stars were observed \citep{murphy2014}.

In this paper we use a hare-and-hounds exercise to investigate the sensitivity of the PM method to various orbits. The individual roles of the hare and the hound are described in Sect.\,\ref{ssec:hare} and \ref{ssec:hound}, respectively. Particular attention has been paid to recovering the orbital parameters for undersampled orbits (Sect.\,\ref{ssec:undersampling}). In addition, we conducted specific experiments to determine the detection limits of the PM method, in terms of both companion mass and maximum orbital period that can be analysed, and the influence of the pulsation properties on these limits (Sect.\,\ref{sec:experiments}). The use of RVs as observational inputs alongside time delays is discussed in Sect.\,\ref{sec:rv} and applied to real \textit{Kepler} data, including a binary system in which both stars pulsate.


\section{Method: a hare-and-hounds exercise}

\subsection{Simulating binary systems (the hare)}
\label{ssec:hare}

\subsubsection{Systems with one pulsator}

One of us (HS) generated a series of synthetic light curves of $\delta$ Sct variables.
We adopted theoretically computed eigenfrequencies of an evolutionary model of a 1.8-M$_{\odot}$ star at the mid-main-sequence stage. We varied the pulsation content by including modes with spherical degrees, $\ell$, up to $\ell=4$. The star was treated as non-rotating, so that the azimuthal orders of each mode have the same (degenerate) frequency. Mode amplitudes were scaled approximately to their disk-integrated intensity, which decreases with increasing $\ell$. The goal was to generate pulsation spectra that are broadly similar to those observed by \textit{Kepler}. Specific examples are given in Sect.\,\ref{sec:experiments}.

The sinusoidal luminosity variations of each mode were added together as a 1500-d time series, roughly coinciding with the duration of the \textit{Kepler} mission. Gaussian random noise of $\sim$0.2\,mmag was added at each time cadence, corresponding to the typical noise per measurement for a 13th-magnitude \textit{Kepler} target. The sampling time interval, $\delta t$, was set to 30\,min, corresponding to \textit{Kepler}'s long-cadence mode. Barycentric corrections to the time stamps were taken into account. See \citet{murphyetal2012b} for details on the corrections for \textit{Kepler}'s orbit.

To simulate a binary system we introduced the delay in arrival time of the light due to the orbital motion by adding a time-dependent term to the flux values:
\begin{equation}
	L(t) = \sum_{j} A_j \cos(2\uppi\nu_j [t - \tau(t)] ),
\label{eq:02}
\end{equation}
where $A_j$ and $\nu_j$ are the amplitude and the frequency of mode $j$.
For a given orbit, the time delay is expressed as a function of the true anomaly, $f$, by
\begin{equation}
	\tau(t) = -{{a_1\sin i}\over{c}} {{1-e^2}\over{1+e\cos f}}\sin(f+\varpi). 
\label{eq:03}
\end{equation}
Here, $a_1\sin i$ denotes the projected semi-major axis, $e$ is the eccentricity, $\varpi$ is the angle between the node and the periastron, and $c$ is the speed of light.
The trigonometric functions of $f$ are expressed in terms of a series expansion of 
the trigonometric functions of the mean anomaly, $l$, by
\begin{equation}
	\cos f = -e + {{2\left(1-e^2\right)}\over{e}} \sum_{n=1}^\infty J_n(ne)\cos nl 
\label{eq:04}
\end{equation}
and
\begin{equation}
	\sin f = 2\sqrt{1-e^2} \sum_{n=1}^\infty J_n'(ne)\sin nl ,
\label{eq:05}
\end{equation}
where $J_n(x)$ denotes the $n^{\rm th}$ Bessel function of the first and integer kind, and $J'_n(x) = {\rm d}J_n(x)/{\rm d}x$. 
Here, the mean anomaly is given as a function of time, $t$, by 
\begin{equation}
	l (t) := 2\uppi\nu_{\rm orb} (t-t_{\rm p}) ,
\label{eq:06}
\end{equation}
where $\nu_{\rm orb}$ and $t_{\rm p}$ denote the orbital frequency and the time of periastron passage, respectively. 
In practice, the infinite sum in equations (\ref{eq:04}) and (\ref{eq:05}) must be truncated; we used 100 orders when injecting orbits into the data.

The light curves were delivered to the hound, but the orbital parameters, $a_1\sin i$, $e$, $\varpi$, $t_{\rm p}$, and $\nu_{\rm orb}$, and the pulsation mode properties were kept secret. Note that by definition the orbital period, $P_{\rm orb}$ and the orbital frequency $\nu_{\rm orb}$ are directly related via
\begin{equation}
	P_{\rm orb} := \frac{1}{\nu_{\rm orb}} 
\label{eq:07}
\end{equation}
and we use these interchangeably.

\subsubsection{Systems with two pulsators}

In some binary systems, both components are observed to pulsate. The mass ratio of these systems can be obtained without dependence on $\sin i$, making them very attractive to study. We refer to these as `PB2' systems in analogy to the SB2 spectroscopic binaries. Their scientific promise motivated us to include PB2s in some of the simulations, with both components being $\delta$\,Sct stars.

\subsection{Analysis of simulated binaries (the hound)}
\label{ssec:hound}

\subsubsection{Obtaining the phased time-delay curve}
\label{sec:method:pm1}

The first task in determining the orbital parameters was to obtain a series of time delay observations. We used the method of \citet{murphyetal2014}, which we summarise here. We began by taking a Fourier transform of the stellar light curve, and selecting the peaks of highest amplitude in the range 5--44\,d$^{-1}$. Peaks below 5\,d$^{-1}$ were not used for two main reasons: firstly, this frequency region is more likely to contain peaks of non-pulsational origin (e.g. due to noise, which could be the red noise of a low-mass companion or instrumental noise), and secondly, peaks that are of pulsational origin are likely to be g\:modes like those found in $\gamma$\,Dor stars, which are not typically suitable for PM analysis (see \citealt{comptonetal2016}). Since such peaks nonetheless contribute to the variance in the data, affecting the uncertainties in the pulsation phases, we attenuated this low frequency region, i.e. the data were high-pass filtered. The corresponding frequency region of 5-d$^{-1}$ width just below the sampling frequency ($\sim$49\,d$^{-1}$) was also ignored, giving the upper limit of 44\,d$^{-1}$. Nyquist aliases were avoided by selecting the peak of highest amplitude among a set of Nyquist ambiguities (see \citealt{murphyetal2012b} for a description).

The frequencies of those highest peaks were determined with the full 1500-d light curve using a non-linear least-squares routine. The light curve was then subdivided into shorter segments, and the phase of the peaks (at fixed frequency) was measured in each segment using a linear least-squares routine. Phase changes were converted into time delays by dividing by the angular frequency of each peak. We calculated a series of weighted mean time delays, weighting by the phase uncertainties for each of the extracted peaks in each of the light-curve segments.

\begin{figure}
\begin{center}
\includegraphics[width=0.48\textwidth]{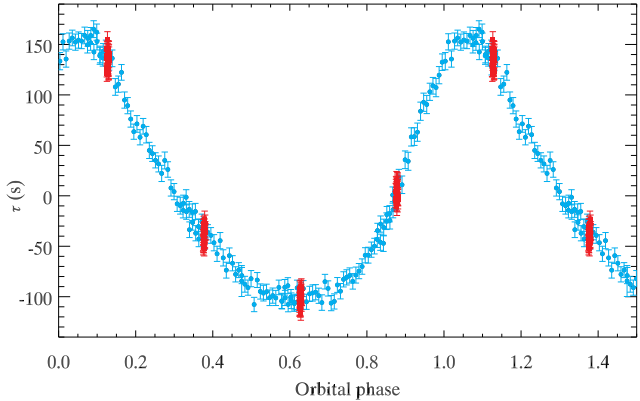}
\caption{Observed time delays at slightly different sampling: at 10.00\,d (red squares) and at 9.92\,d (blue circles). The orbital period is 40.00\,d, so a non-integer segment size provides much better phase coverage.}
\label{fig:phase_coverage}
\end{center}
\end{figure}

The Fourier transform of the weighted mean time delays has a peak at the orbital frequency, from which the orbital period was obtained. The time delays were then folded on this frequency. After phase folding, it became evident that the default segment size (10\,d) was inappropriate for some of the simulated binaries because they had orbital periods that were close to integer multiples of 10\,d. This leads to poor phase coverage in the time-delay curve and poorly constrained orbital parameters (Fig.\,\ref{fig:phase_coverage}), so for those time-delay curves the process was repeated with modified segment size.

\subsubsection{Determining the orbital parameters}

After producing a time-delay curve for each star, we applied the PM2 method \citep{murphy&shibahashi2015} to obtain analytical values of the other orbital parameters: the projected light travel time across the orbit, $a_1 \sin i$, eccentricity, $e$, argument of periastron, $\varpi$, and phase of periastron, $\phi_{\rm p}$. The analytical approach calculates approximate values for these parameters, but does not constrain their uncertainties unless the parameter space surrounding those values is explored. We therefore implemented an MCMC routine.

The MCMC approach used a Metropolis-Hastings algorithm \citep{metropolisetal1953,hastings1970} with symmetric proposal distributions based on gaussian-distributed random numbers. The standard deviations of the proposal distributions were initially chosen based on the analytical orbital parameters already determined. Smaller proposal steps were made for more eccentric orbits, where $\varpi$ and $\phi_{\rm p}$ are better defined. The process is not adaptive in real-time, but we made trial runs to ensure the proposal acceptance rate lay in an appropriate range, taken to be between 0.15 and 0.50. Proposals incorporated steps in all five orbital parameters ($P_{\rm orb}$, $a_1 \sin i$, $e$, $\varpi$, and $\phi_{\rm p}$) simultaneously. As is standard, we accepted the proposed state if its likelihood ($L_{\rm prop.} = {\rm exp}[{\chi}^2/2]$) was greater than that of the current state. If the current state had the greater likelihood, then we evaluated the likelihood ratio ($L_{\rm prop.} / L_{\rm current}$) against a uniform random number between 0 and 1, and accepted the proposed state if the likelihood ratio exceeded this number. We manually checked the posterior distributions to ensure the parameter space had been explored and sampled appropriately.

The final values of the orbital parameters were determined as the medians of the marginalised posteriors. An alternative option would be to take the mean value in the chain for each parameter, or the mode of the histograms of the marginalised posteriors. The decision of which to take is arbitrary when the parameter space is well sampled, the posterior distribution is gaussian, and the number of proposals is sufficiently high to avoid small-number statistics. In all cases, data are discarded when collapsing the marginalised posteriors to single values. The uncertainties on the orbital parameters were determined as the points corresponding to 0.159 and 0.841 in the cumulative distribution of the marginalised posteriors, which therefore bracket the central 68.2\:per\:cent of the data.

\subsection{Sampling of the orbit}

\label{ssec:undersampling}

The segment size needs to be long enough to give adequate frequency resolution in each segment, but that can mean that the orbit is not well-sampled, particularly near periastron. Severe smearing of $a_1 \sin i / c$ and $e$ occurs when the time delays are undersampled. In this subsection we derive the coefficients necessary to compensate for undersampling, such that the true orbital elements can be recovered. The results from the hare-and-hounds exercise are discussed in Sect.\,\ref{sec:experiments}.

\subsubsection{Instantaneous sampling: a Fourier series expression of the time delay}

\begin{figure}
\begin{center}
\includegraphics[width=\linewidth, angle=0]{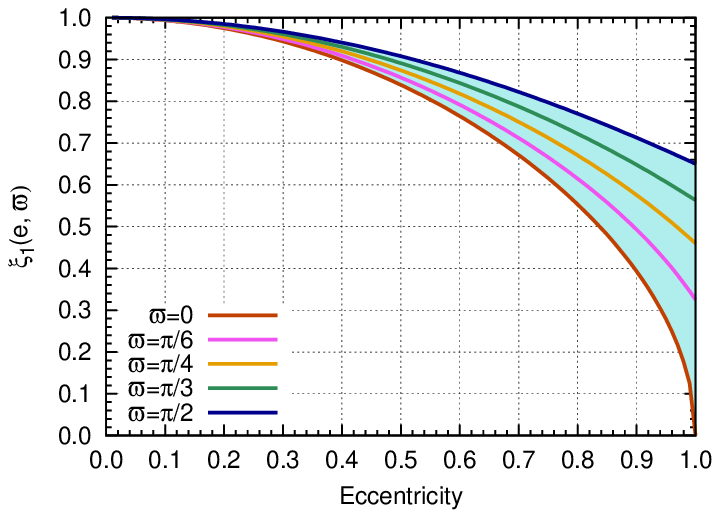} 
\includegraphics[width=\linewidth, angle=0]{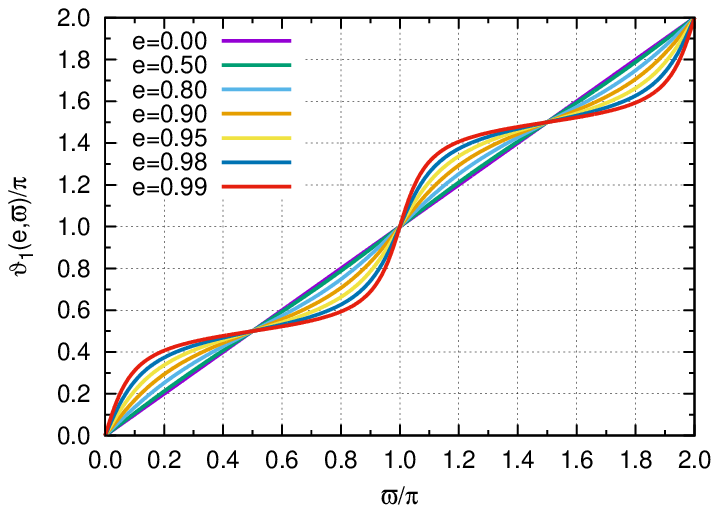}
\caption{ Top: The coefficient $\xi_1(e,\varpi)$ as a function of eccentricity $e$ for different values of $\varpi$. The band shows the full range of $\varpi$ from $0$ to $2\uppi$. For $e \ll 1$, $\xi_1(e, \varpi) \sim 1$, while its $\varpi$-dependence grows with the increase of $e$. Bottom: The phase $\vartheta_1(e,\varpi)$ as a function of $\varpi$ for different values of $e$. The $e$-dependence grows rapidly with $e$. Note that both functions are shown for the case of $n=1$.
}
\label{fig:xi_1}
\end{center}
\end{figure}

With the help of equations (\ref{eq:03})--(\ref{eq:06}), the time delay is written as a function of time
\begin{equation}
	\tau (t) 
	=
	-{{a_1\sin i}\over{c}}
	\sum_{n=1}^\infty \xi_n
	\sin [ 
	2\uppi n\nu_{\rm orb}
	 (t-t_{\rm p}) + \vartheta_n]
	-\tau_0 ,
\label{eq:08}
\end{equation}
where
\begin{equation}
	\xi_n(e,\varpi) 
	:=
	{{2}\over{n}} J_n'(ne) 
	\sqrt{1-\left[1- \left({ {\sqrt{1-e^2}}\over{e}} {{J_n(ne)}\over{J_n'(ne)}} \right)^2 \right] \cos^2\varpi},
\label{eq:09}
\end{equation}
\begin{equation}
	\vartheta_n(e,\varpi) :=
	\left\{\begin{array}{ll} 
		\arctan\left[{{e}\over{\sqrt{1-e^2}}} {{J_n'(ne)}\over{J_n(ne)}}\tan\varpi\right] & \mbox{if}\ 0\leq \varpi  < \frac{\uppi}{2} \\
		\arctan\left[{{e}\over{\sqrt{1-e^2}}} {{J_n'(ne)}\over{J_n(ne)}}\tan\varpi\right] + \uppi & \mbox{if}\ \frac{\uppi}{2} \leq \varpi < \frac{3\uppi}{2} \\
		\arctan\left[{{e}\over{\sqrt{1-e^2}}} {{J_n'(ne)}\over{J_n(ne)}}\tan\varpi\right] + 2\uppi & \mbox{if}\ \frac{3\uppi}{2} \leq \varpi < 2\uppi 
	\end{array}\right.
\label{eq:10}
\end{equation}
and
\begin{equation}
	\tau_0(e,\varpi) := -{{a_1\sin i}\over{c}}\sum_{n=1}^\infty \xi_n\sin (-
	2\uppi n\nu_{\rm orb}
	t_{\rm p}+\vartheta_n) .
\label{eq:11}
\end{equation}
Here $\arctan(x)$ returns the principal value of the inverse tangent of $x$. See \cite{shibahashietal2015} for further details. As seen in the top panel of Fig.\,\ref{fig:xi_1} for the case of $n=1$, the $\varpi$-dependence of $\xi_n(e, \varpi)$ is weak for $e \ll 1$, but it becomes important with the increase of $e$. Similarly, the $e$-dependence of $\vartheta_n(e, \varpi)$ is weak for $e \ll 1$, but it dramatically grows as $e$ approaches unity, as shown in the bottom panel of Fig.\,\ref{fig:xi_1}.

\subsubsection{Discrete sampling: the dependence of smearing on the segment size}

Equation (\ref{eq:08}) indicates that the orbital elements can be determined from the time delay in the case of instantaneous sampling. However, in the present method, we divide the observational time span into non-overlapping segments of size $\Delta t$, and deal with the time delay averaged over the segment.\footnote{The segment size $\Delta t$ is of the order of 10\,d and should not be confused with the \textit{Kepler} long-cadence sampling rate $\delta t$ (=30\,min).}
This causes the maxima and minima of the time delay curve to become less sharp, hence the orbital elements deduced from it systematically deviate from the true values.

The time delay averaged over the $i^{\rm th}$ time segment \mbox{$[t_i - \Delta t/2, t_i+\Delta t/2]$} is deduced as 
\begin{equation}
	\bar{\tau} (t_i) 
	=
	-{{a_1\sin i}\over{c}}
	\sum_{n=1}^\infty \xi_n
	{\rm sinc} \left( n\uppi\eta \right) \sin [2\uppi n\nu_{\rm orb} (t_i-t_{\rm p}) + \vartheta_n]
	-\tau_0,
\label{eq:12}
\end{equation}
where
\begin{equation}
	\eta := \nu_{\rm orb}\Delta t
\label{eq:17}
\end{equation}
is the reciprocal of the number of segments per orbit and is not usually an integer.
The sinc terms in the right-hand side of equation (\ref{eq:12}) show the smearing effect due to undersampling. Of course, in the limit of $\Delta t =0$, equation (\ref{eq:12}) tends to equation (\ref{eq:08}).


\subsubsection{Treatment of undersampled time delays in the MCMC software}
\label{sssec:e_correction}

In order to correct for smearing due to undersampling, we fit the functional form of equation (\ref{eq:12}) directly to the observed time delays. The resulting orbital parameters are then equivalent to those that would be determined if the sampling were instantaneous.

\begin{figure*}
\begin{center}
\includegraphics[width=0.475\linewidth, angle=0]{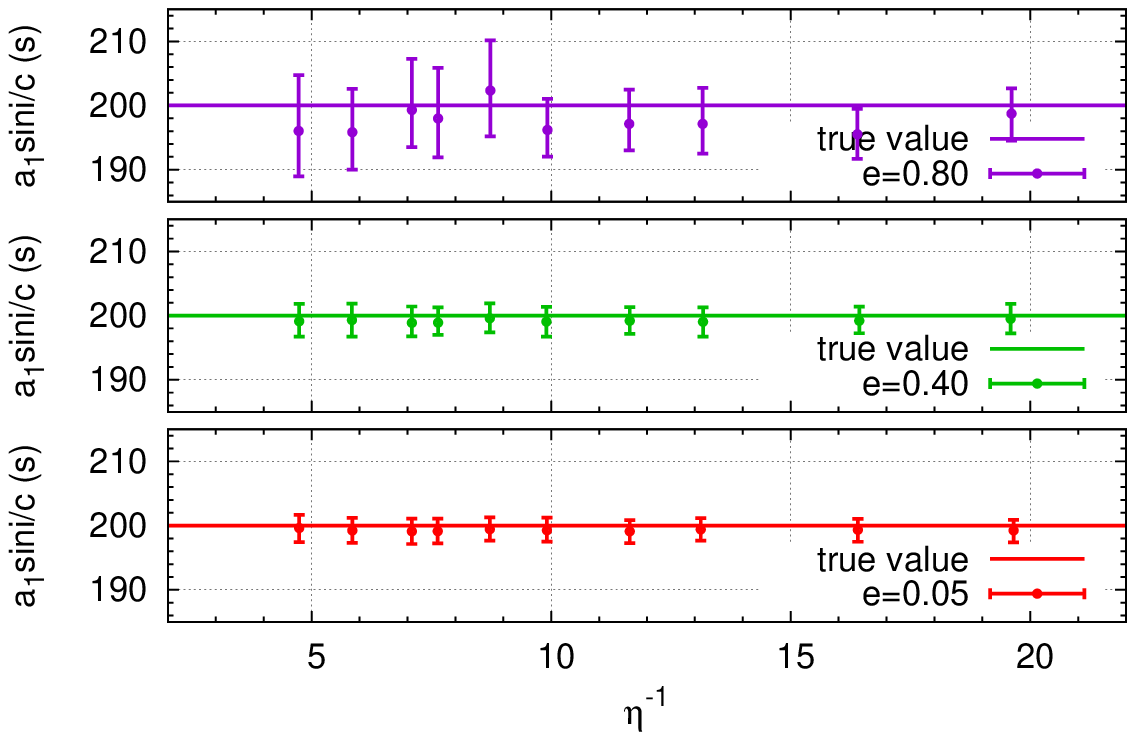}\hspace{5mm}
\includegraphics[width=0.475\linewidth, angle=0]{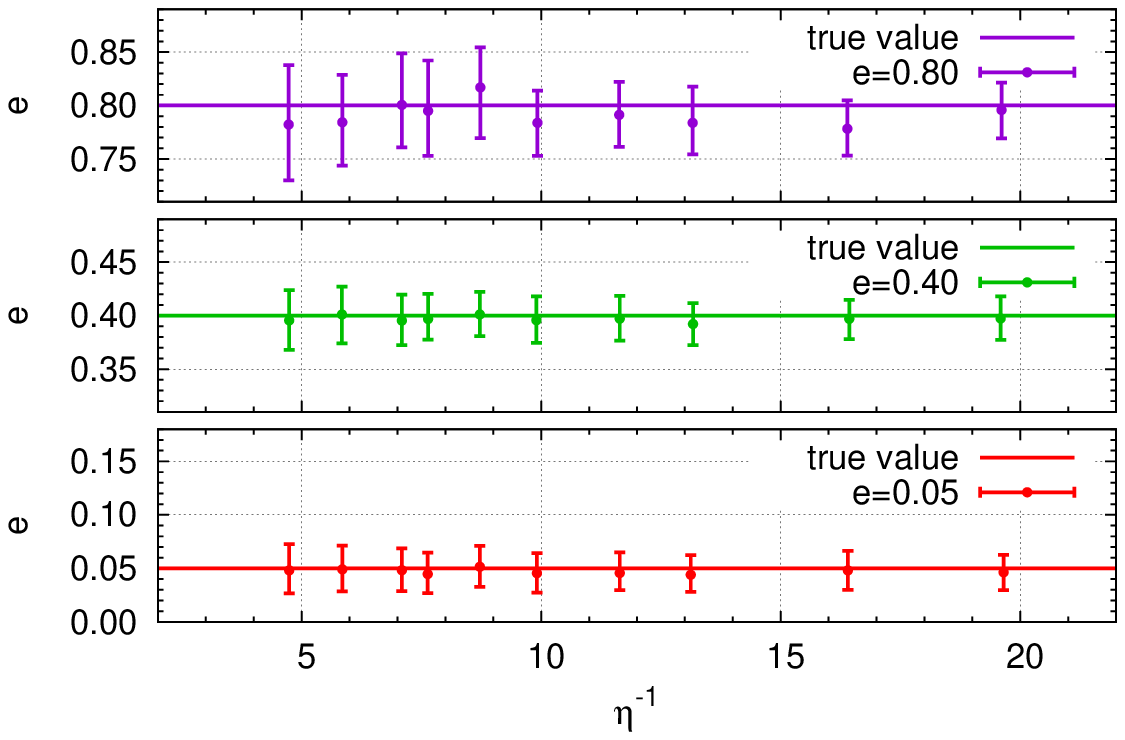}
\includegraphics[width=0.475\linewidth, angle=0]{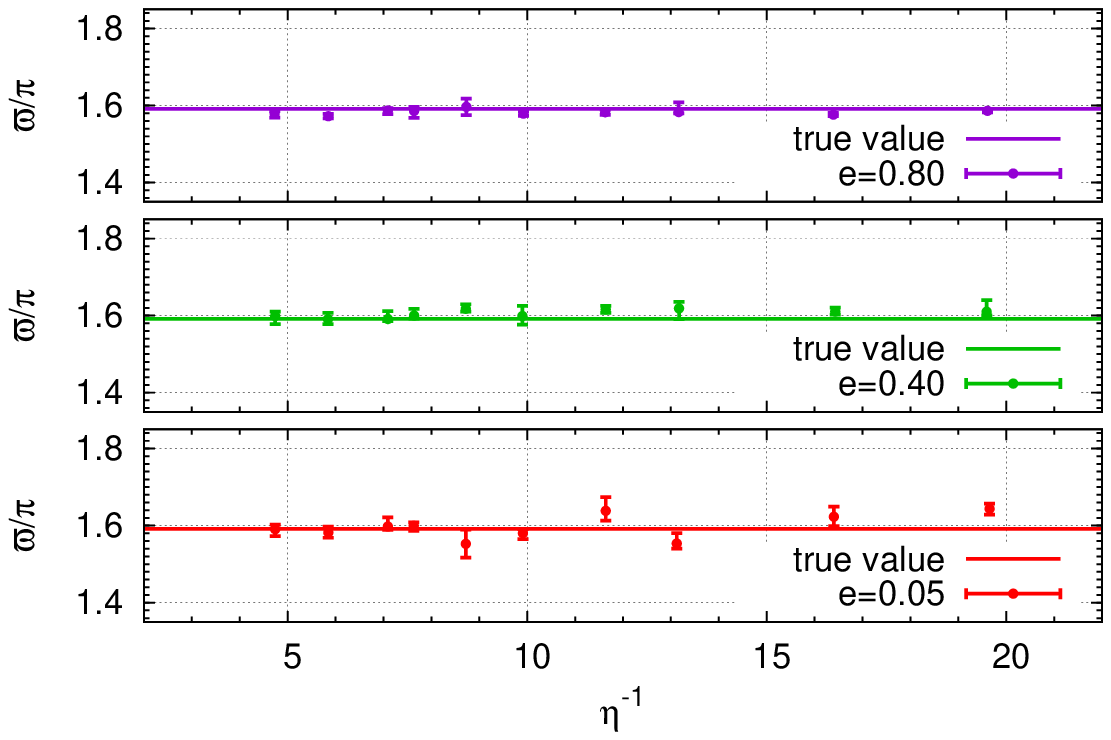}\hspace{5mm}
\includegraphics[width=0.475\linewidth, angle=0]{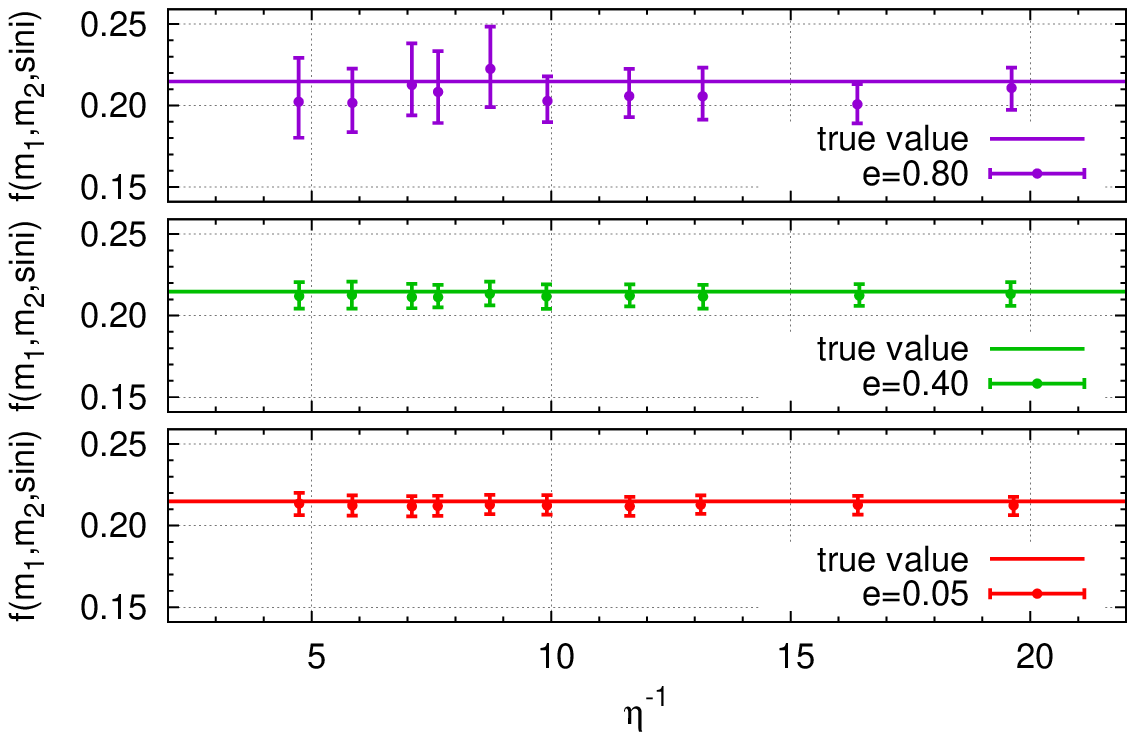}
\caption{The effect of undersampling on the MCMC outputs (filled circles), for orbits with eccentricities of $e=0.05, 0.4$, and 0.8. The abscissa of each panel is $1/\eta = (\nu_{\rm orb}\Delta t)^{-1}$. Clockwise from top left, results are shown for the projected semi-major axis $a_1\sin i/c$, eccentricity $e$, the mass function and the argument of periastron $\varpi$. The true values are shown by the horizontal lines.
}
\label{fig:07}
\end{center}
\end{figure*}

We verified this with numerical simulations analysed with different sampling rates (Fig.\,\ref{fig:07}). The simulated orbital elements were: $P_{\rm orb}=200.0$\,d, $a_1\sin i/c=200.0$\,s, $\varpi=5.0$\,rad, and three different values of $e$ were used: 0.05, 0.4, and 0.8.

The upper left panel shows the projected semi-major axis, $a_1\sin i/c$. The input value is clearly reproduced at the $1\sigma$ level, with no dependence on the segment size used or the eccentricity. This value of $a_1\sin i/c$ is later used to calculate the mass function:
\begin{eqnarray}
	f(m_1,m_2,\sin i) &:=& {{(m_2\sin i)^3}\over{(m_1+m_2)^2}}
\label{eq:35}
	\\
	&=&
	{{4\uppi^2c^3}\over{G}} \nu_{\rm orb}^2\left( {{a_1\sin i}\over{c}} \right)^3,
\label{eq:36}
\end{eqnarray}
where $G$ is the gravitational constant and $m_2$ denotes the mass of the companion.
Thus, with a suitable assumption of the primary mass $m_1$, the mass of the companion can be found.

The upper right panel of Fig.\,\ref{fig:07} shows that the eccentricity, $e$, is also well reproduced, even when the smearing effect is strong. The argument of periastron, $\varpi$, in the lower left panel, has larger uncertainties when the eccentricity is small and periastron becomes hard to identify, but is recovered satisfactorily even for $e=0.05$. The projected semi-major axis is determined less precisely -- although still within 1$\sigma$ -- when the eccentricity is high.

\subsubsection{Choice of the segment size}

Given that the orbital parameters are well-recovered even when the number of segments per orbit is large, how large should segment sizes be? The answer reflects a trade-off between the desire to detect short-period orbits and the need to resolve modes with nearby frequencies. If many low degree, high order acoustic modes are excited in a target star, the frequencies of $\ell=0$ modes with a certain radial order 
and those of $\ell=2$ modes with a radial order lower by one 
are quite close.\footnote{This is the small separation, $\delta \nu$, in solar-like oscillators.} The situation worsens with higher $\ell$, as we will demonstrate in Sect.\,\ref{sssec:pulsation_properties}, and further deteriorates if rotational splittings are included. In order to resolve such small frequency separations in a dense frequency spectrum, the segment size should be longer than the reciprocal of that frequency separation. Other factors to consider are the increase in computation time for processing more time delay measurements versus the improvement in precision of the orbital parameters when $\eta$ is smaller.


\subsubsection{Binning the observations}
The increase in computation time when $\eta$ is small can be alleviated by binning the observations after the time delays have been folded on the orbital period. As long as the number of bins remains adequate for determining the eccentricity, computation time is heavily reduced by taking a weighted mean in each phase bin and scaling the error bars accordingly, with no detriment to the uncertainties on the orbital parameters. A contextualised example will be given in Sect.\,\ref{sec:rv}.

\section{Results from the hare-and-hound experiments}
\label{sec:experiments}

\subsection{Detection limit of the PM method in finding companions}

One of the goals of the hare-and-hounds exercise was to establish the limit down to which the PM method can detect a companion. Here we give a breakdown of the factors that influence that limit for a companion of a given mass, and predict the detection limits of the method as applied to \textit{Kepler} data.

\subsubsection{Dependence on eccentricity}

\begin{figure}
\begin{center}
\includegraphics[width=0.45\textwidth]{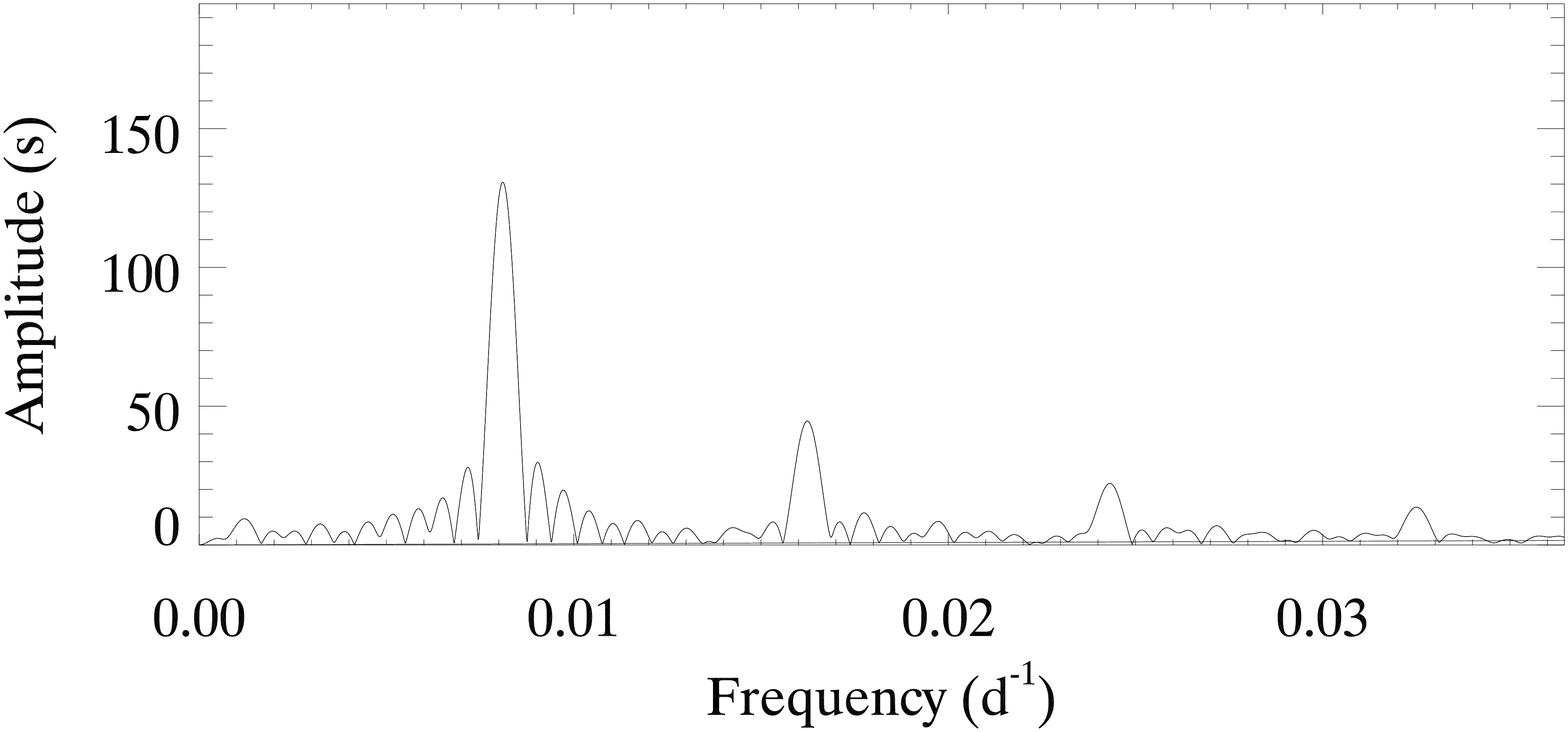}
\includegraphics[width=0.45\textwidth]{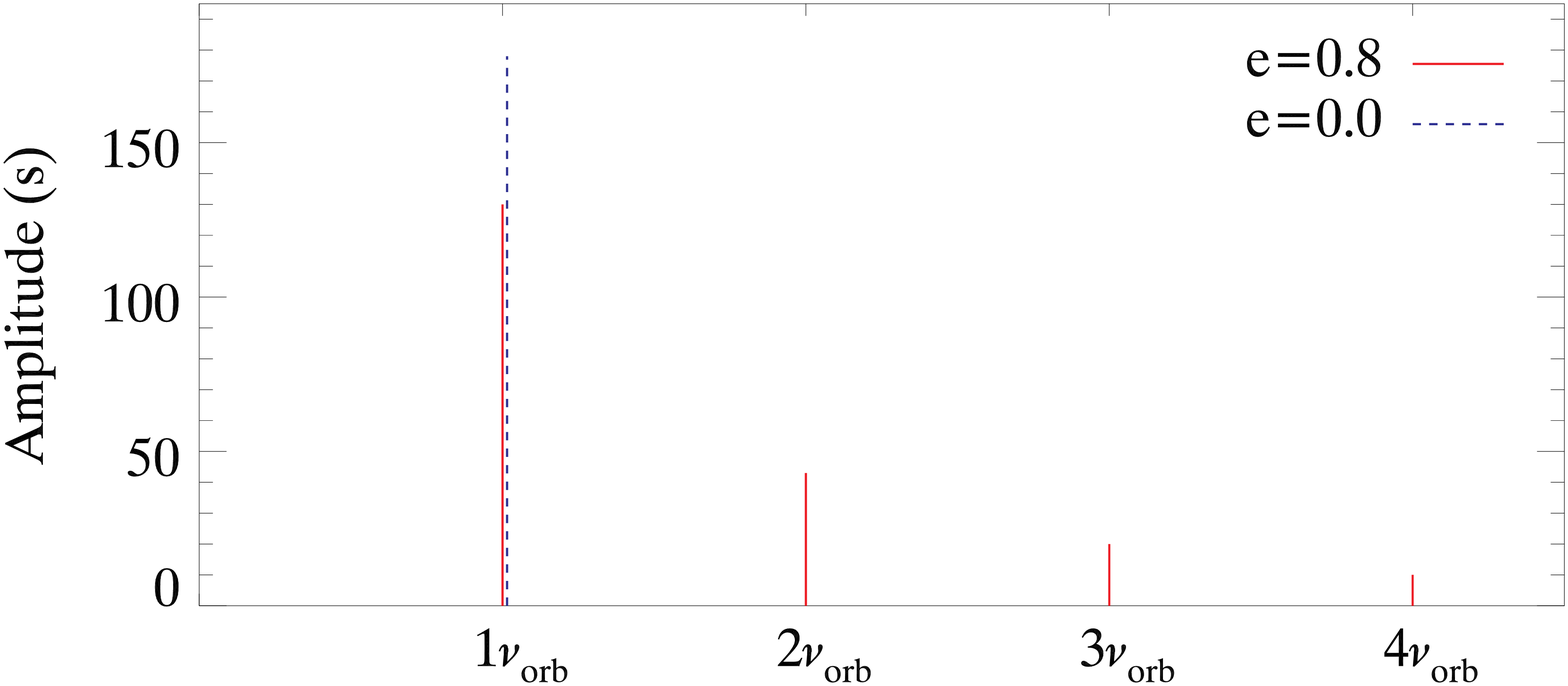}
\includegraphics[width=0.45\textwidth]{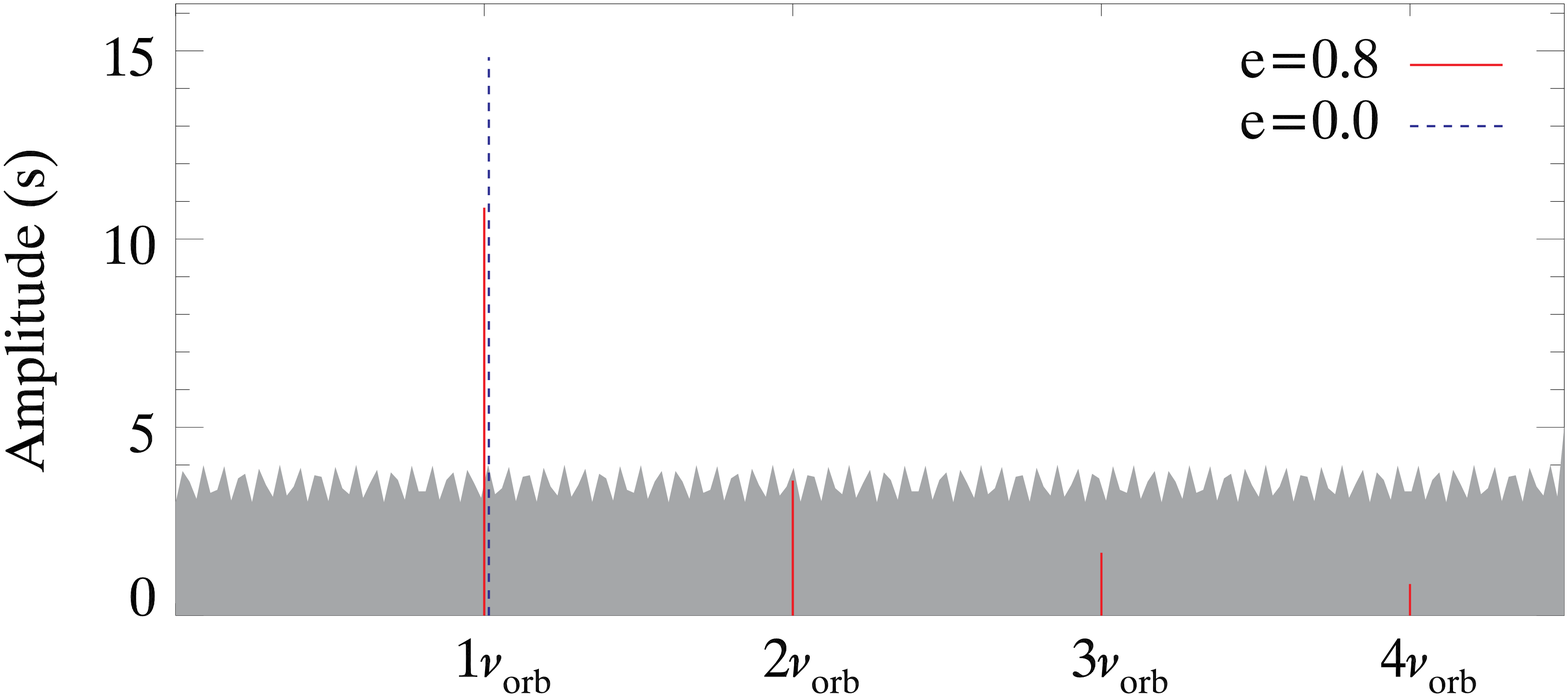}
\caption{Top: Fourier transform of observed time delays for an eccentric orbit ($e=0.8$), showing many harmonics of the orbital frequency. Middle: schematic representation of the top panel, with a similar orbit but with zero eccentricity shown for comparison. Bottom: schematic representation of a simulation with a lower companion mass to assess detection limits. The typical noise level of around 3\,s is shown, which leads to the harmonics of the eccentric orbit being hidden. The signal-to-noise ratio is higher for the circular orbit, thus circular orbits are easier to detect.}
\label{fig:detection_eccentricity}
\end{center}
\end{figure}

A Fourier transform of the time delays for an eccentric orbit features a Fourier series at the orbital frequency \citep[see][especially their figure 21]{murphy&shibahashi2015}. The phases of the harmonics contain information on the value of $\varpi$, while their amplitudes depend on the value of $e$: the higher the eccentricity, the more the amplitude is concentrated into harmonics at the expense of the peak at the orbital frequency. If all other parameters are held fixed, the drop in amplitude of the peak at the orbital frequency gives it a lower significance and makes highly eccentric orbits harder to detect (Fig.\,\ref{fig:detection_eccentricity}).

\subsubsection{Dependence on pulsation properties}
\label{sssec:pulsation_properties}

\citet{comptonetal2016} investigated the suitability of the PM method for different classes of pulsating stars. The pulsation properties are the biggest influence on the detectability of binarity: the frequencies of the oscillations, their signal-to-noise ratios and their separation in frequency all play a role.

The frequencies of the oscillations determine the accuracy of the clock. When the ratio of $a_1 \sin i / c$ to the pulsation period ($1/\nu$) is large, the oscillations function as a better clock and there is greater sensitivity to the orbital variation. This is illustrated in figure\:4 of \citet{comptonetal2016}, alongside the effect of the signal-to-noise of the oscillation mode. Only monoperiodic stars were investigated there; here we present some examples for a multiperiodic oscillator, which allows us to discuss the issue of mode crowding.

\begin{figure*}
\begin{center}
\includegraphics[width=0.49\textwidth]{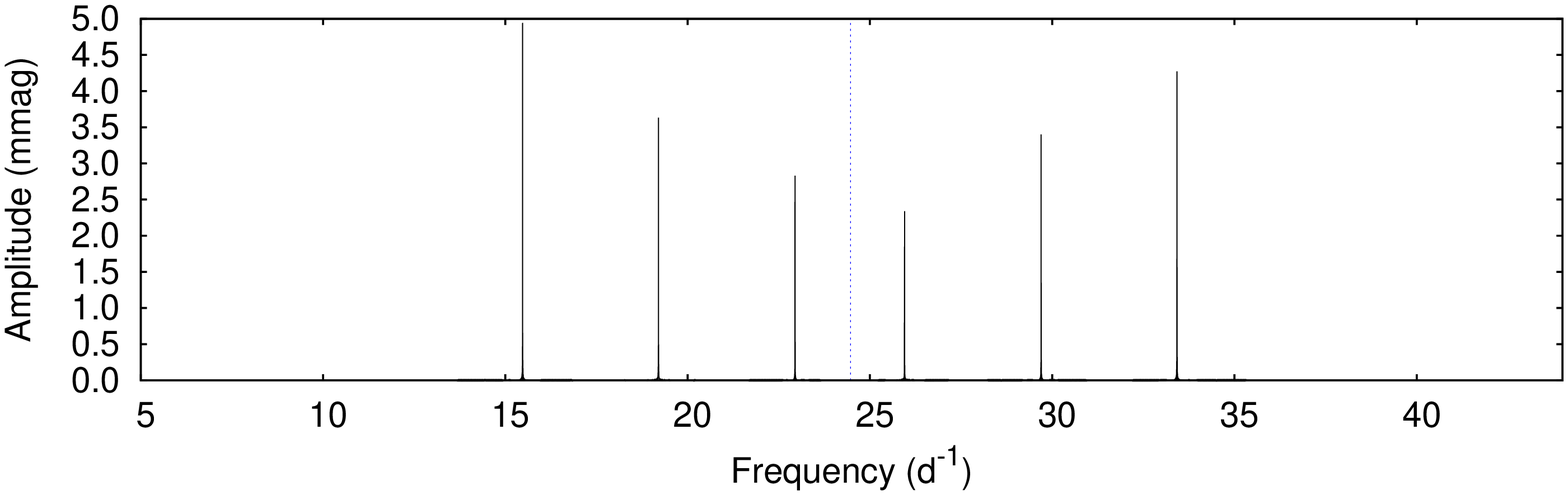}\includegraphics[width=0.49\textwidth]{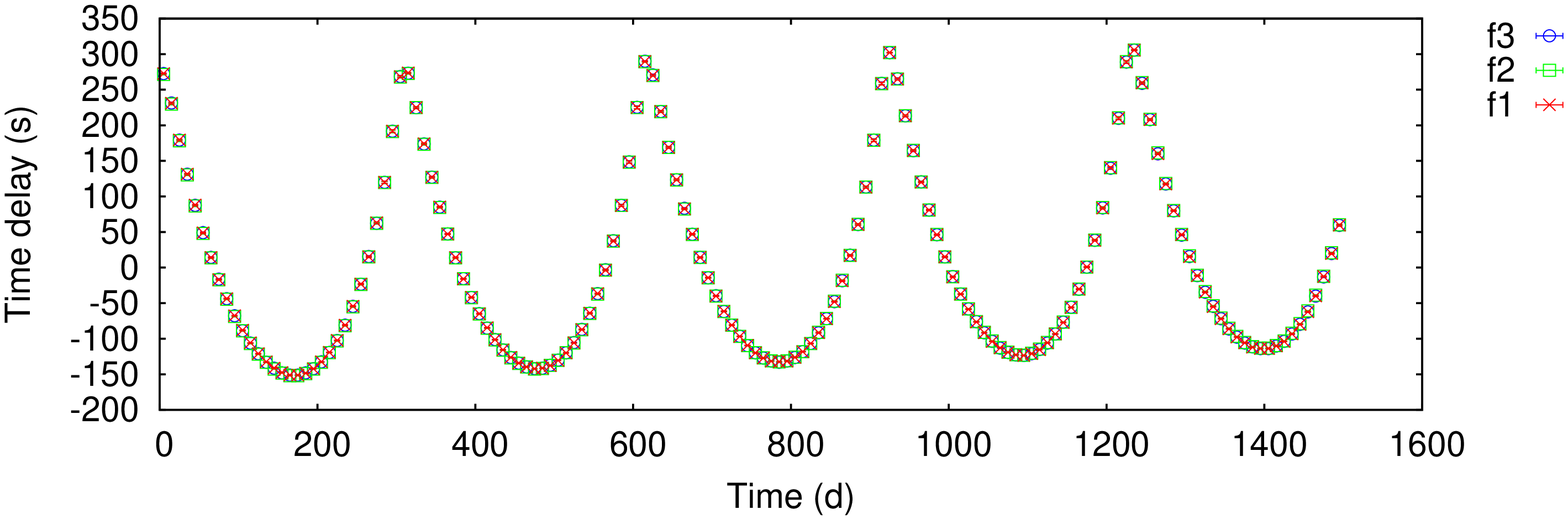}
\includegraphics[width=0.49\textwidth]{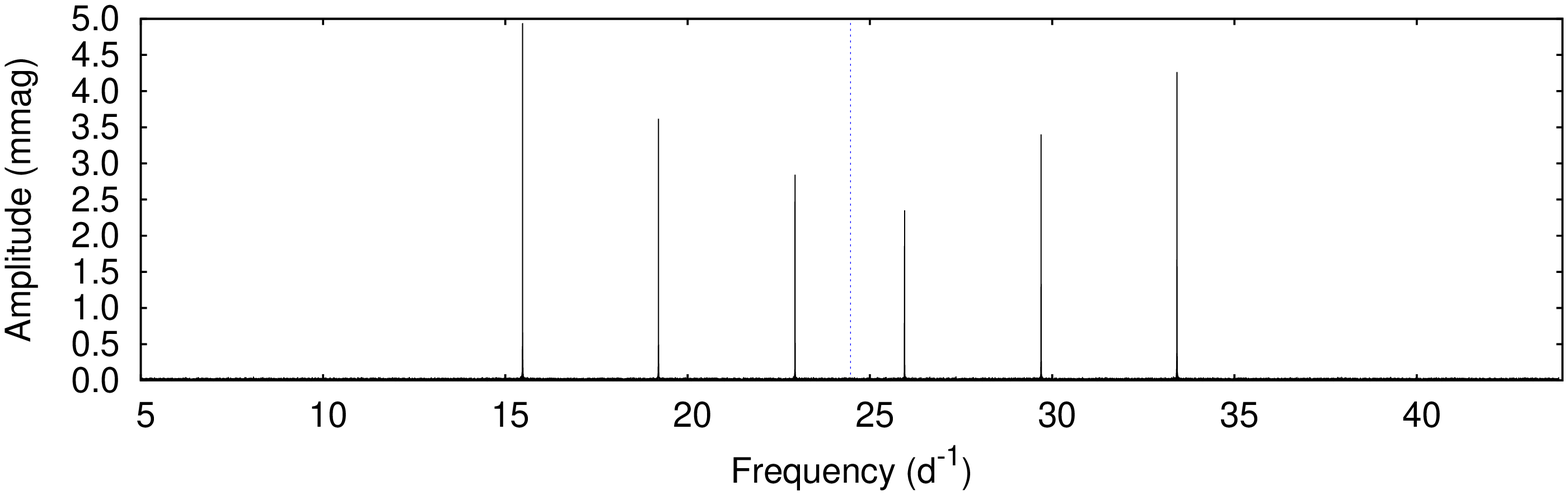}\includegraphics[width=0.49\textwidth]{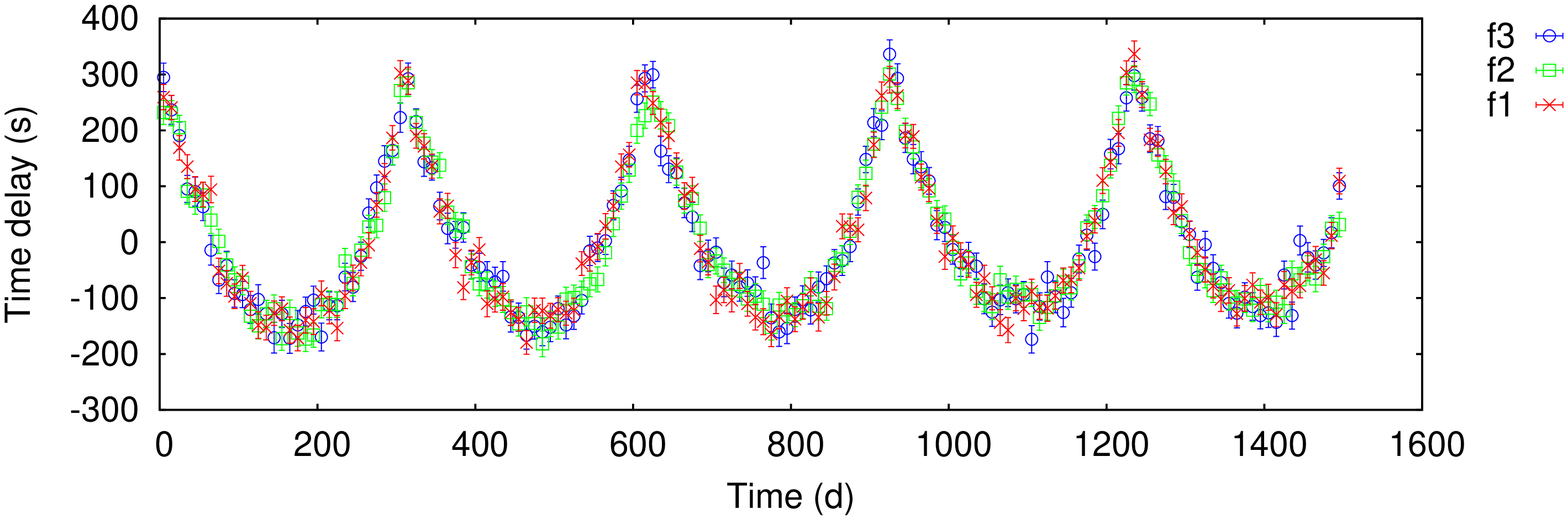}
\includegraphics[width=0.49\textwidth]{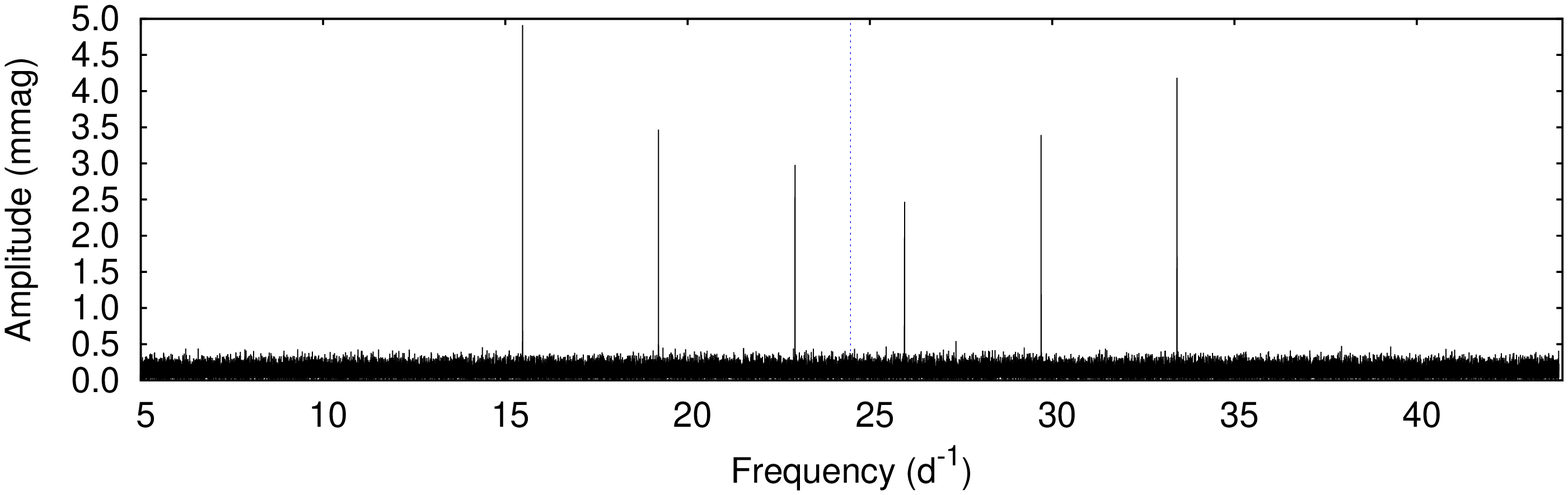}\includegraphics[width=0.49\textwidth]{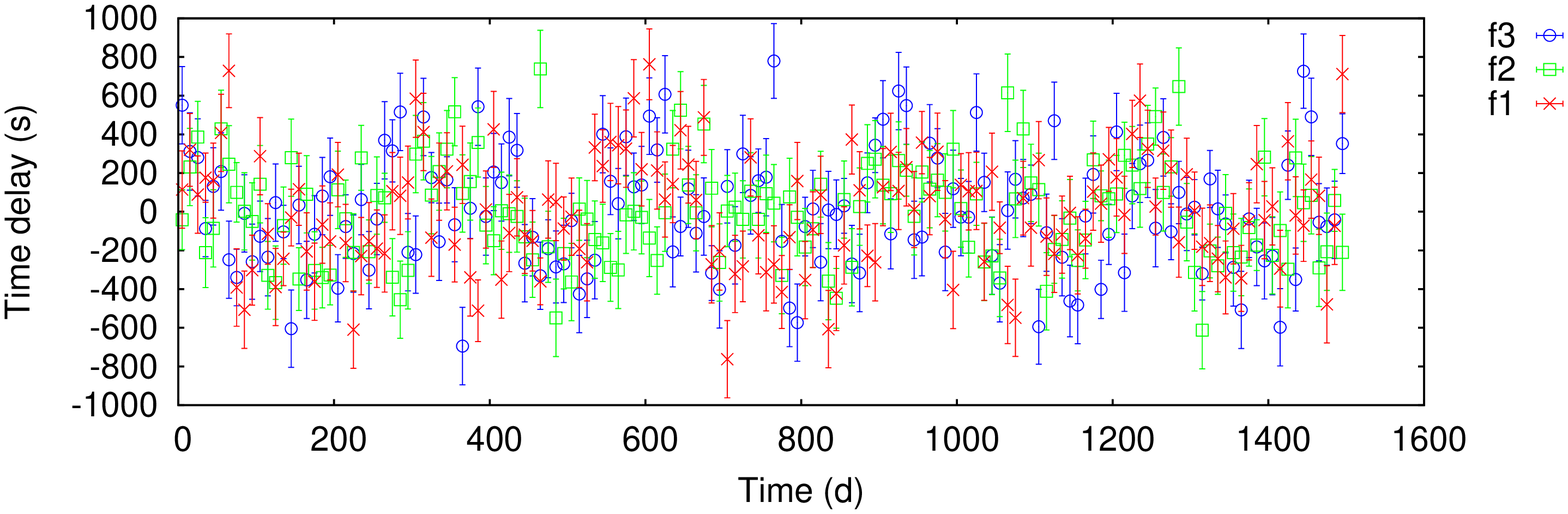}
\caption{Pulsation spectra with oscillation signal-to-noise ratios decreasing downwards, and their corresponding time delay curves. The labels $f_i$ on the time delay curves (right) correspond to the pulsations (left) in order of decreasing amplitude (1=highest). The dashed blue line in the left panels is the Nyquist frequency; the peaks above the Nyquist frequency are all aliases in this case.}
\label{fig:oscillation_noise}
\end{center}
\end{figure*}

\begin{figure*}
\begin{center}
\includegraphics[width=0.49\textwidth]{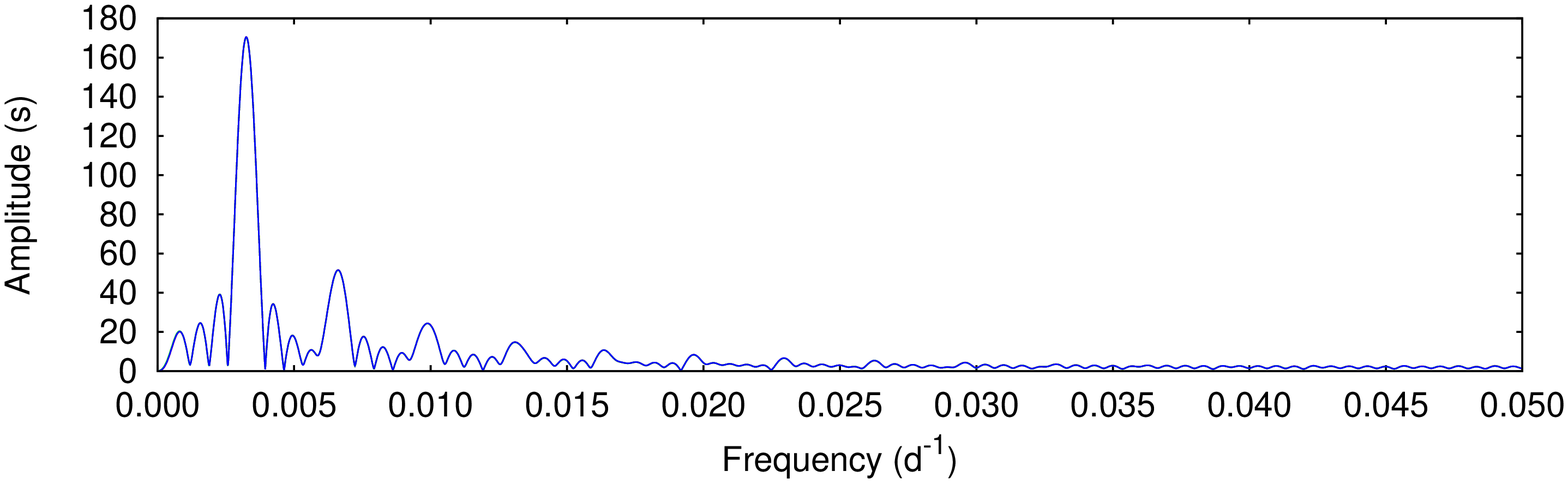}\includegraphics[width=0.49\textwidth]{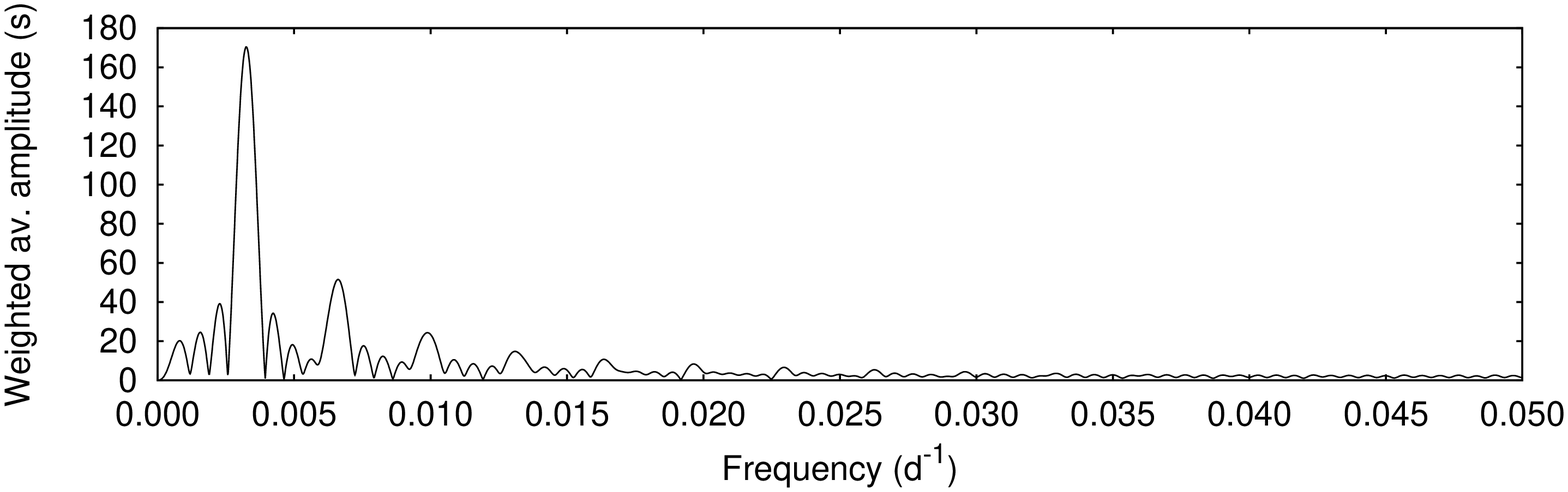}
\includegraphics[width=0.49\textwidth]{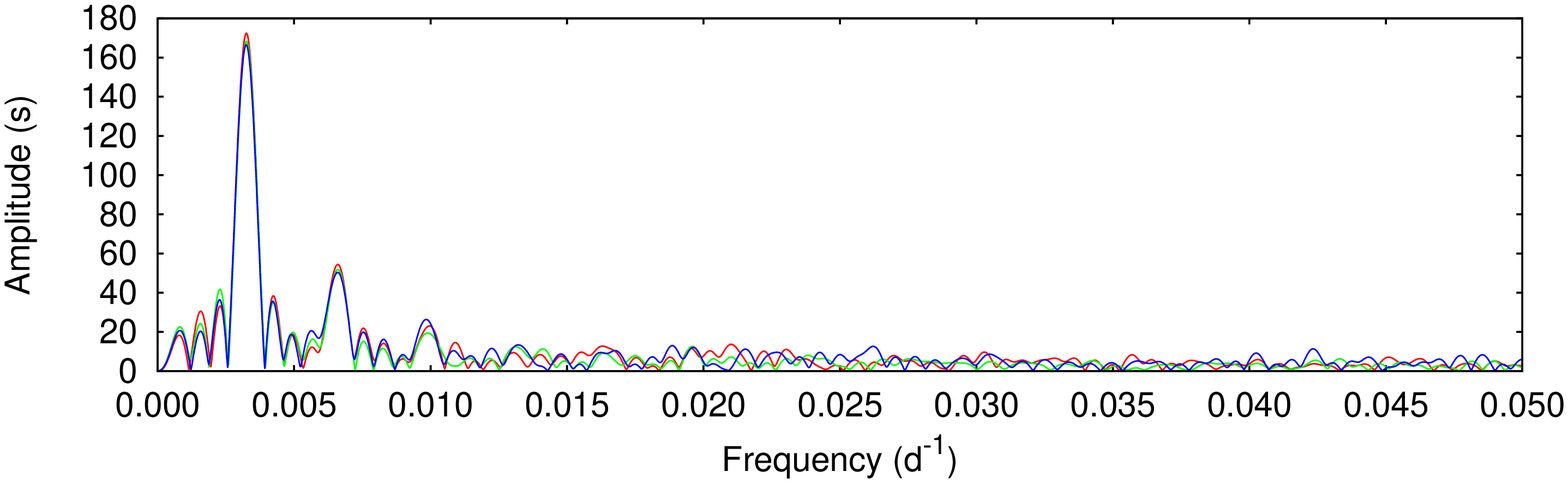}\includegraphics[width=0.49\textwidth]{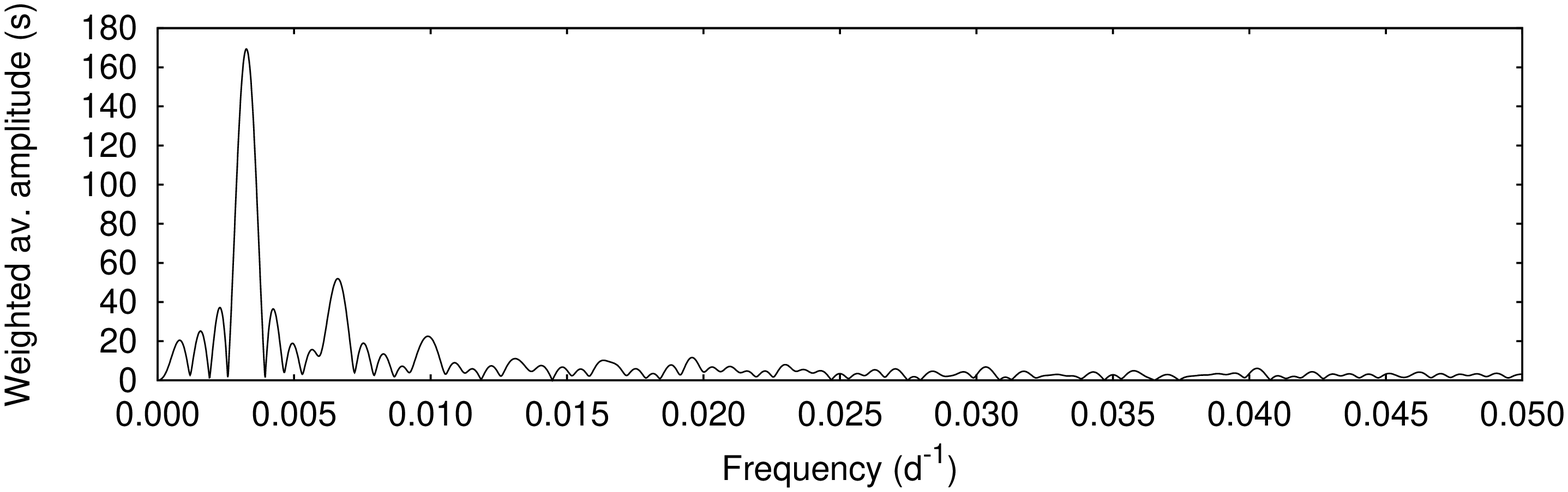}
\includegraphics[width=0.49\textwidth]{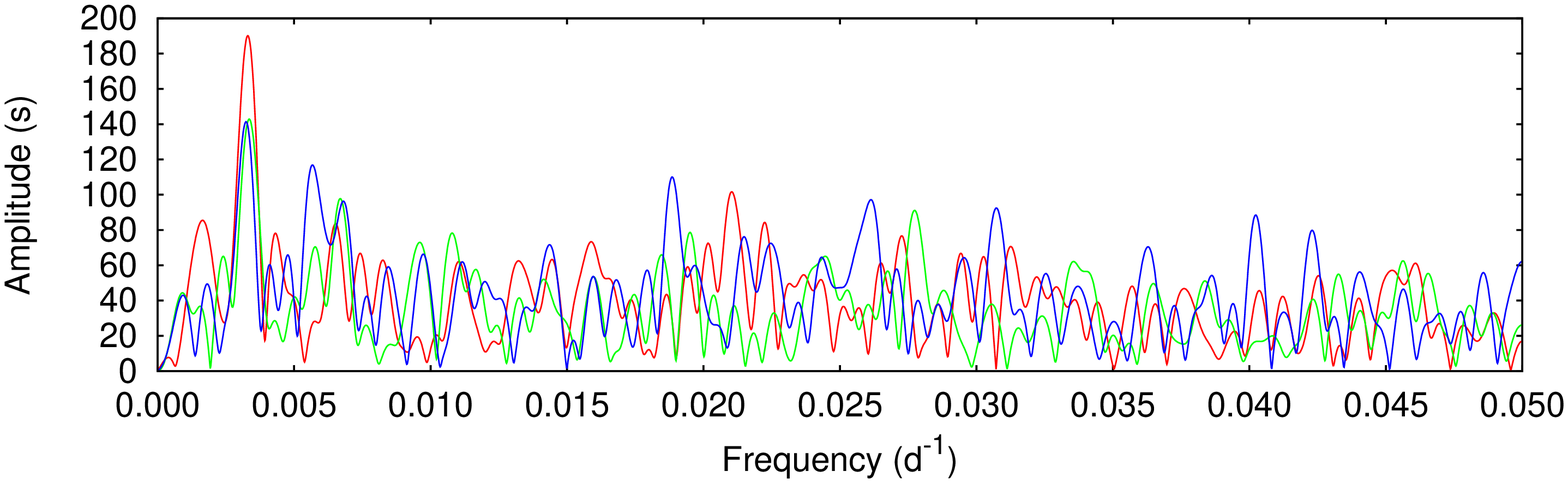}\includegraphics[width=0.49\textwidth]{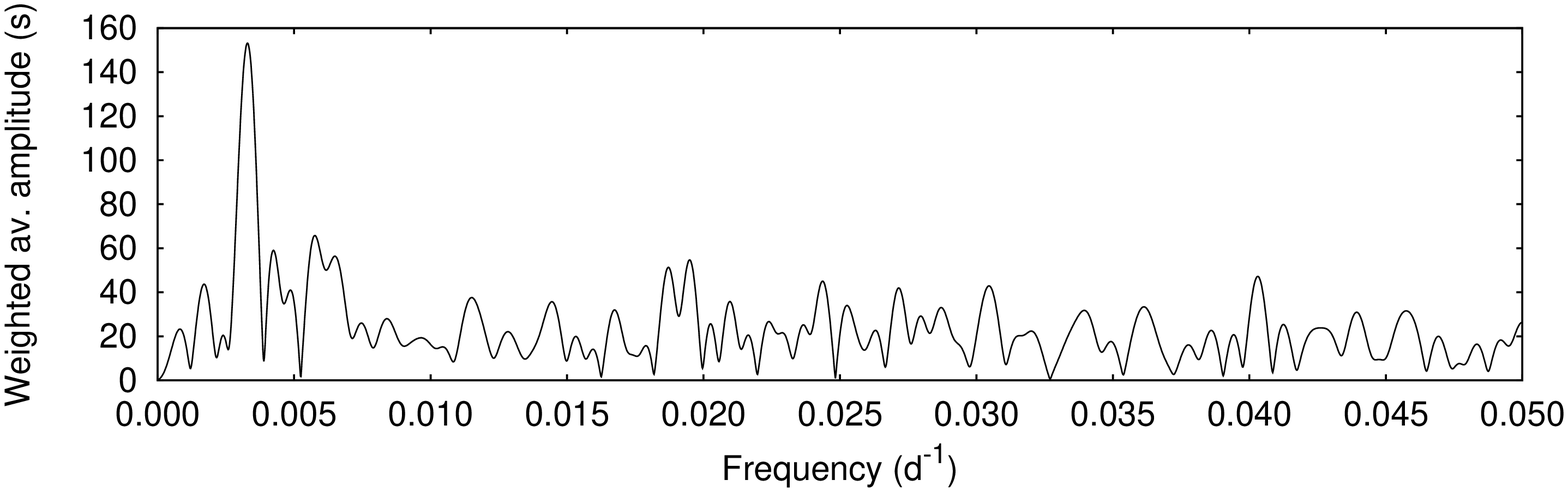}
\caption{Fourier transforms of the time delays of individual modes (left) and the weighted average time delays (right) for the simulations in Fig.\,\ref{fig:oscillation_noise}. White noise in the light curve leads to white noise in the time delays. Notice that, in the bottom panels, the orbital harmonics due to the eccentricity are no longer obvious.}
\label{fig:oscillation_noise_ft}
\end{center}
\end{figure*}

We investigated the influence of the signal-to-noise of the oscillations by simulating three identical binary systems whose pulsation spectra were identical apart from different levels of Gaussian-distributed random noise: (i) no noise, (ii) 0.2\,mmag per point, and (iii) 2.0\,mmag per point. The oscillation frequencies were those of the first, second and third radial overtone modes of a 1.8-M$_{\odot}$ star at the mid main-sequence stage. They were therefore well separated in frequency, allowing us to assess the effect of mode signal-to-noise without mode crowding being a factor. The Fourier transforms of each case are shown in Fig.\,\ref{fig:oscillation_noise} for a 1500-d light curve, along with the corresponding time delays. The addition of white noise to the light curve leads to an increase in scatter in the time delays. Fig.\,\ref{fig:oscillation_noise_ft} shows that the increased scatter has no frequency dependence (i.e. the time delay noise is also white). Importantly, the extra noise has buried the orbital harmonics, which will lead to a very poorly constrained orbital solution.

\begin{figure}
\begin{center}
\includegraphics[width=0.485\textwidth]{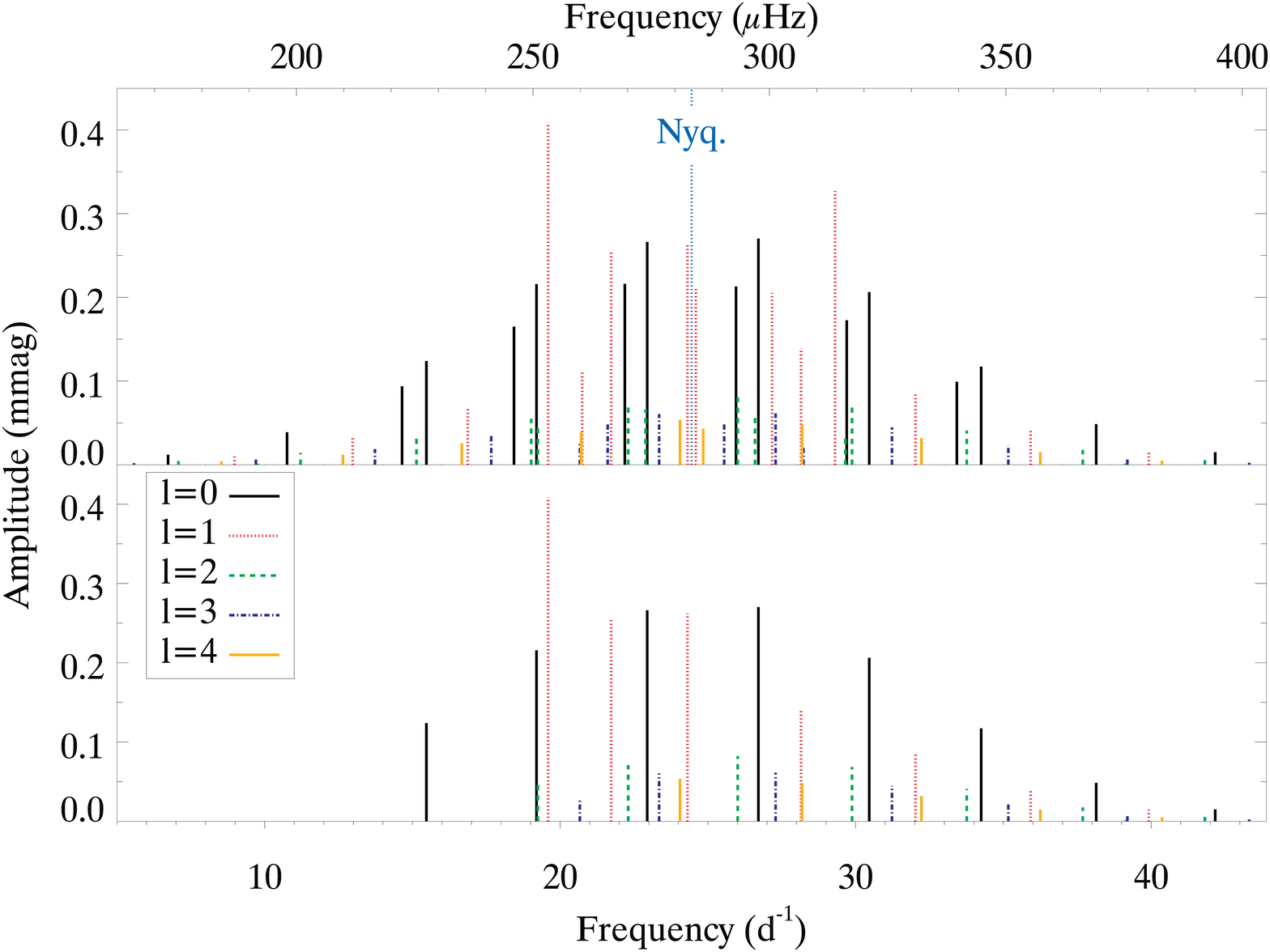}
\caption{Schematic diagram showing the constituents of the `typical' pulsation spectrum. The bottom panel shows the input frequencies and their angular degrees, while the top panel also includes the Nyquist aliases for 30-min sampling, with the Nyquist frequency drawn and labelled. The quadrupole ($\ell=2$) modes cluster tightly around the radial ($\ell=0$) modes, and are not resolved in 10-d segments.}
\label{fig:frequencies_and_ell}
\end{center}
\end{figure}

\begin{figure*}
\begin{center}
\includegraphics[width=0.49\textwidth]{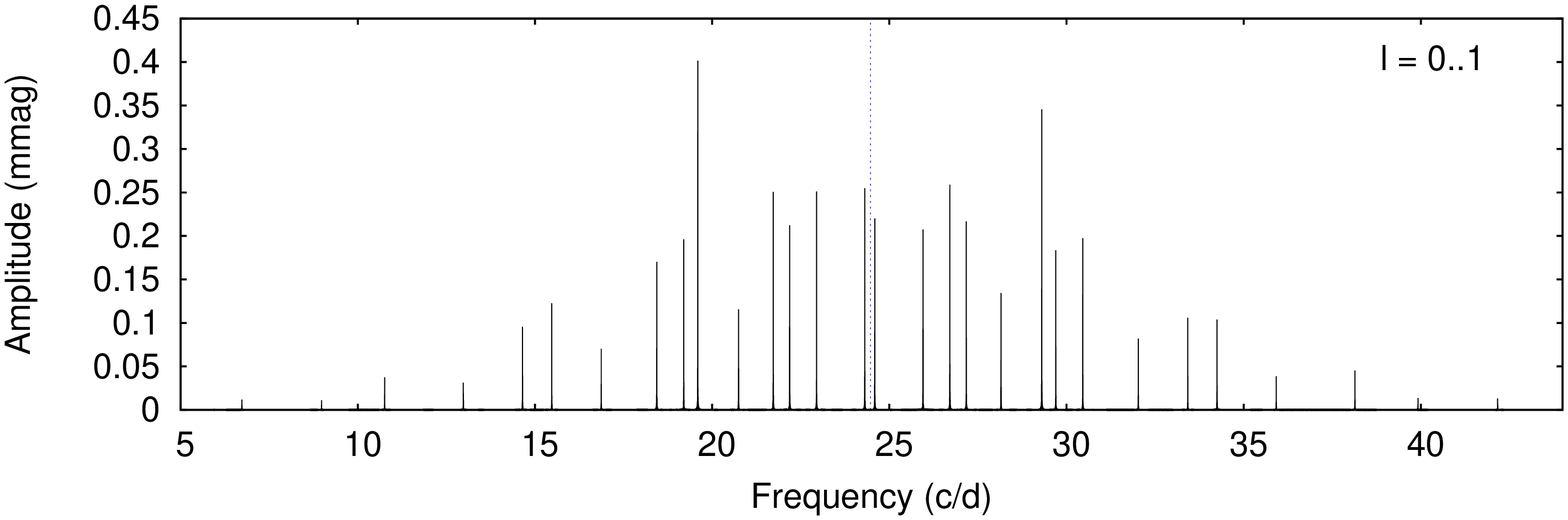}\includegraphics[width=0.49\textwidth]{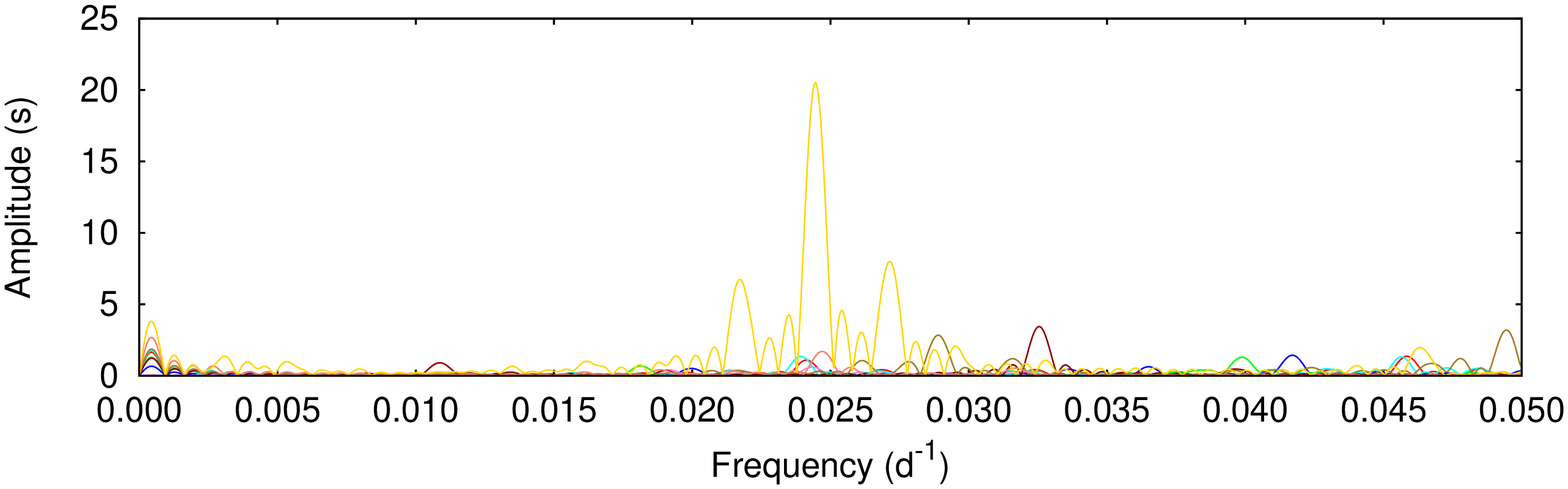}
\includegraphics[width=0.49\textwidth]{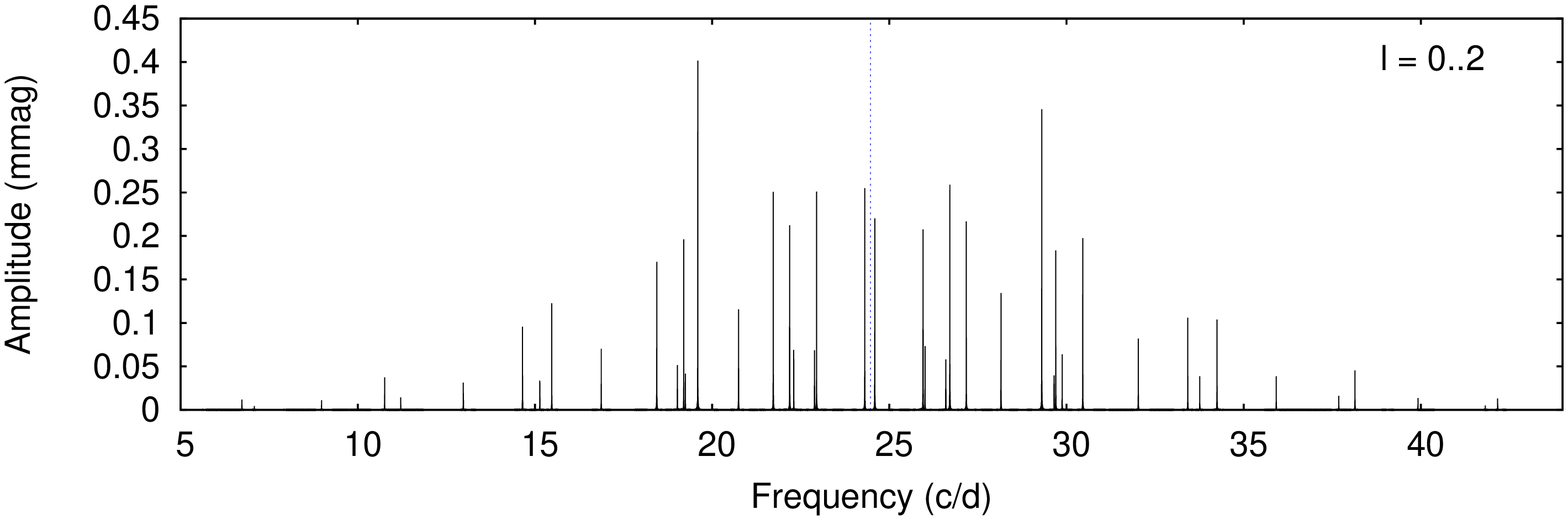}\includegraphics[width=0.49\textwidth]{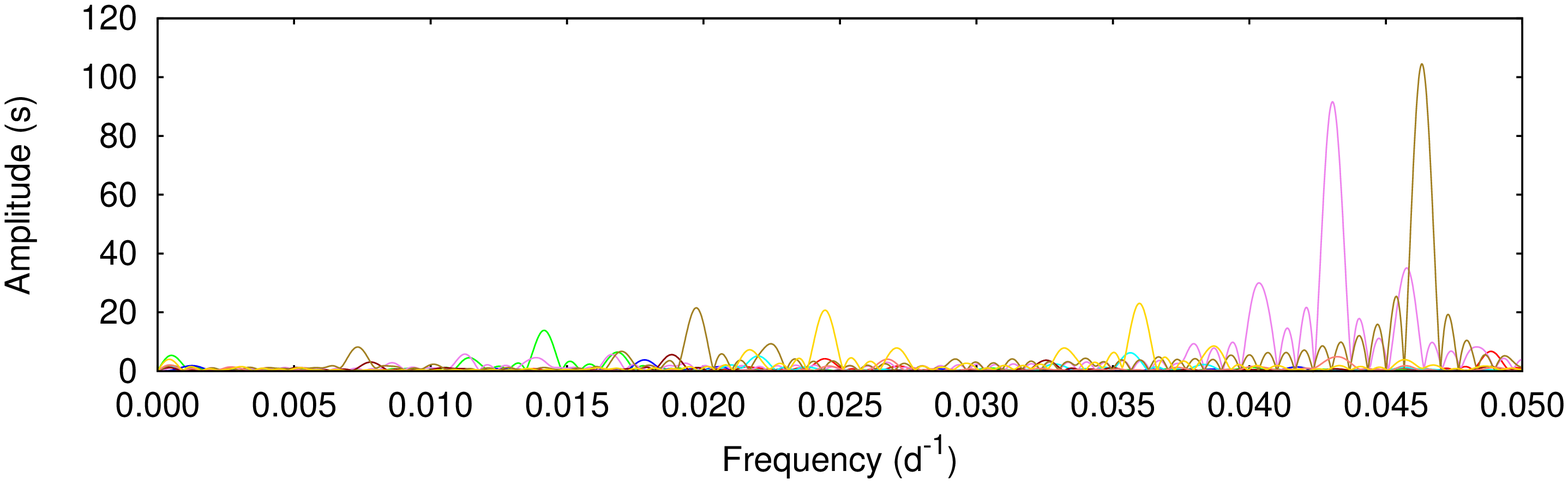}
\includegraphics[width=0.49\textwidth]{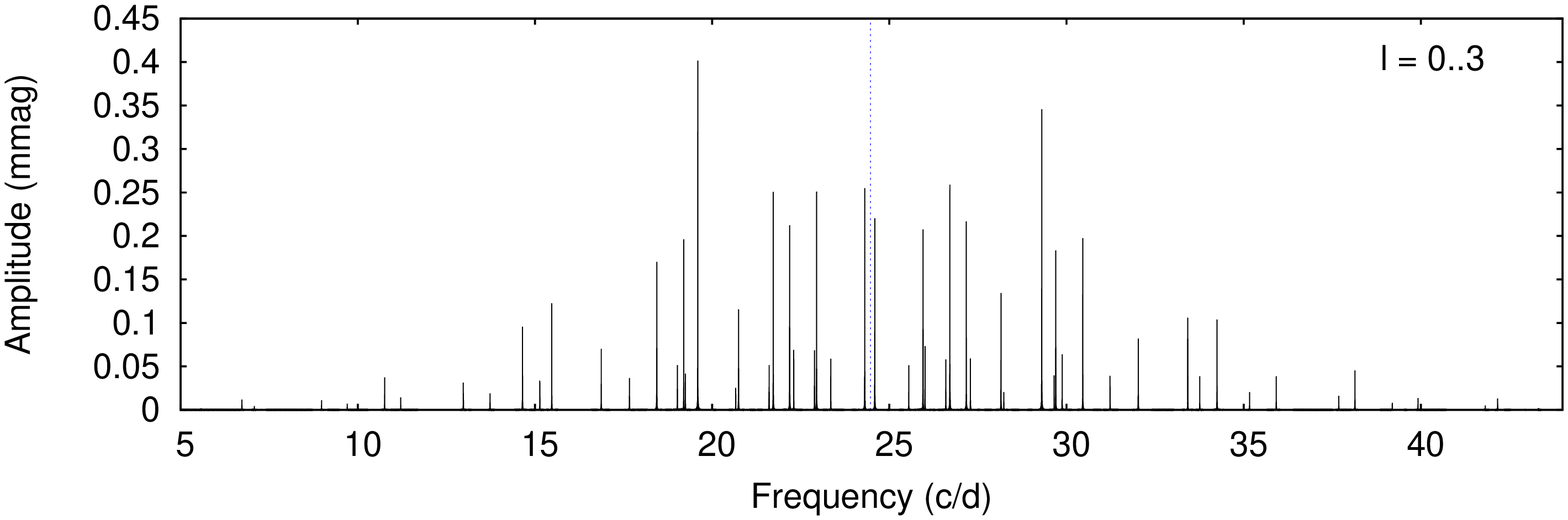}\includegraphics[width=0.49\textwidth]{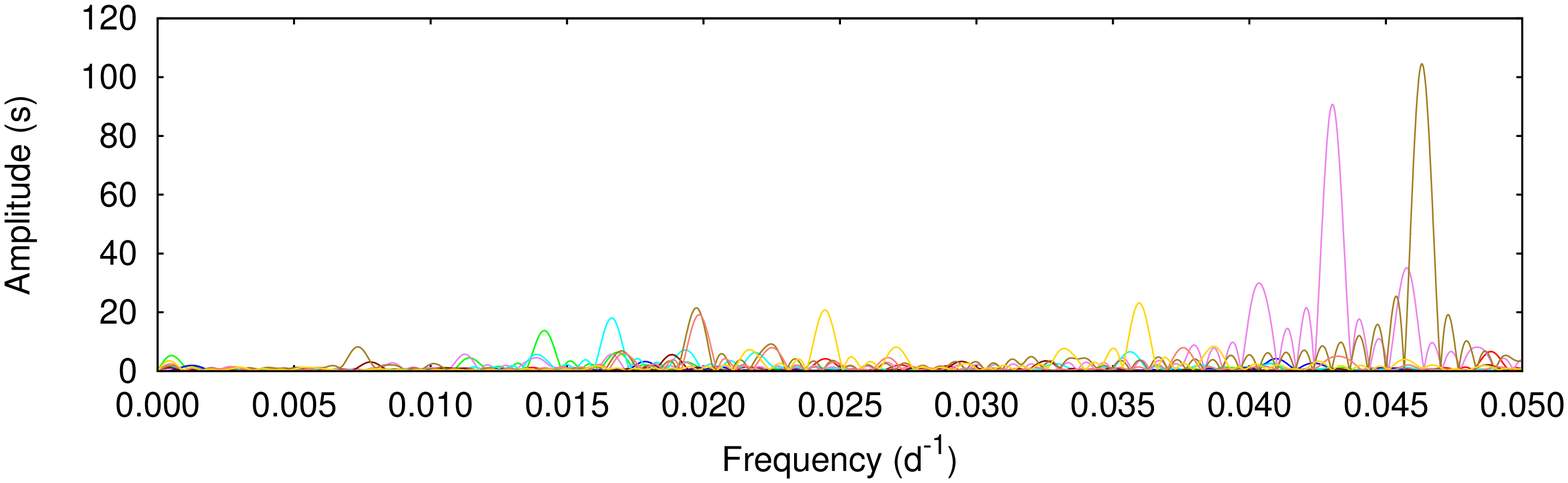}
\includegraphics[width=0.49\textwidth]{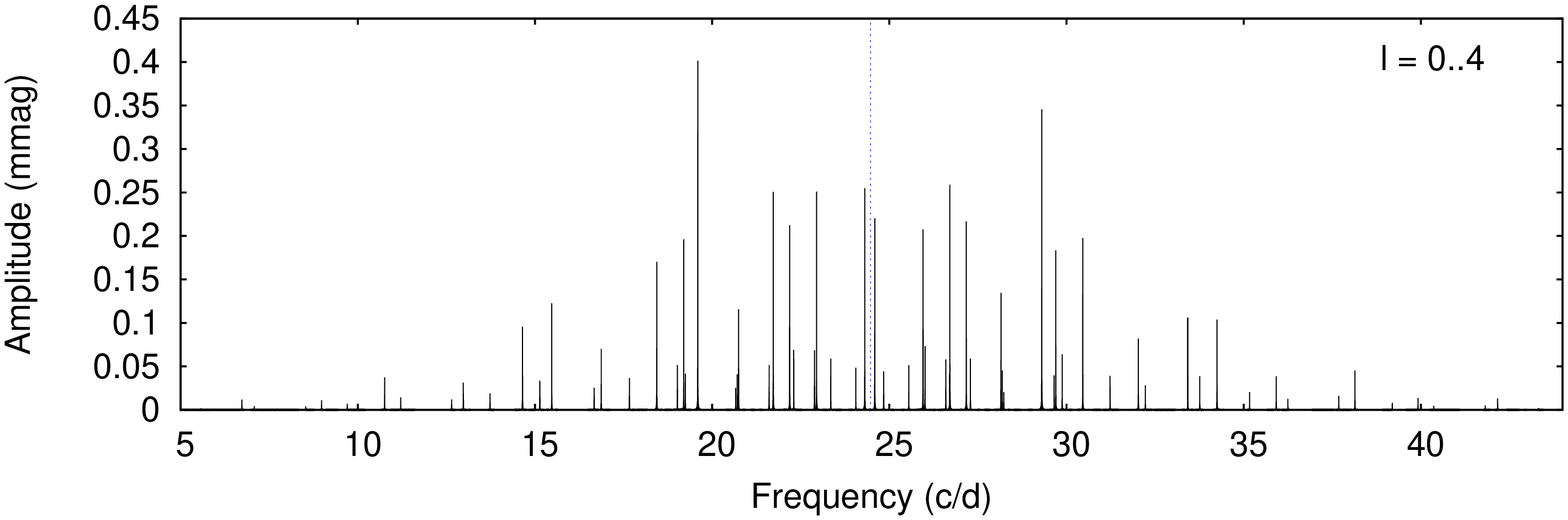}\includegraphics[width=0.49\textwidth]{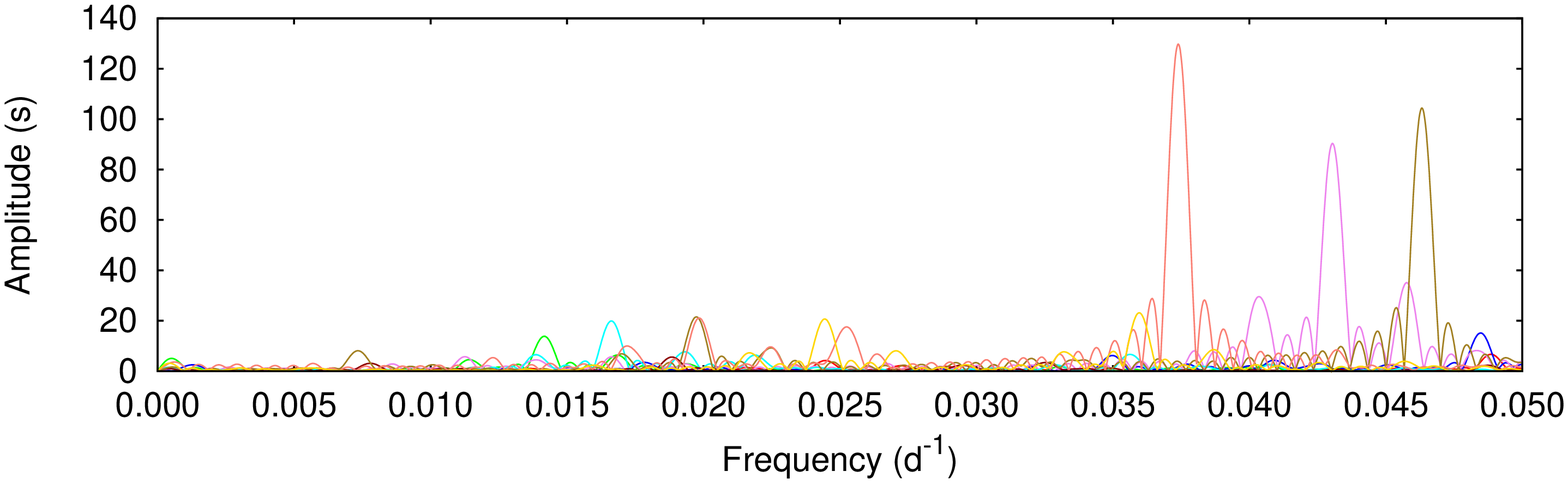}
\caption{Simulations showing that the inclusion of extra modes leads to some closely spaced peaks that are unresolved in short ($\sim$10-d) segments, causing strong and non-white noise at the beat frequencies of the unresolved modes.}
\label{fig:mode_crowding}
\end{center}
\end{figure*}

To investigate the effect of mode crowding, we simulated four additional light curves with no noise. Mode density was controlled by including modes of progressively higher spherical degree, from $\ell=1$ in the simplest case to $\ell=4$ in the densest case. Their frequencies and amplitudes are shown in Fig.\,\ref{fig:frequencies_and_ell}. Since the effect of close mode frequencies is to produce spurious variations in the time delays at their beat frequency \citep{murphyetal2014}, we expect the noise to be non-white. We therefore simulated these light curves without any injected binarity, allowing the spurious frequencies to be seen more clearly. The pulsation spectra and the spurious peaks in the time delays are shown in Fig.\,\ref{fig:mode_crowding}. As more modes are included, the number of spurious peaks grows. In particular, we note that in the models the $\ell=2$ modes tend to cluster around the radial modes, which is also observed in real data \citep{bregeretal2009}. The example shown here is simplistic, because the star is non-rotating and all $2\ell + 1$ azimuthal orders have the same frequency. Spurious peaks can be more burdensome in practice.

The amplitudes of the spurious peaks can easily exceed the amplitudes of the orbital frequency, even when the weighted average is taken, so their treatment is important. One option is to set the weight of each affected mode to zero when calculating the weighted average, but then the binary signal belonging to that mode is also lost. This option is particularly undesirable if the mode has a high amplitude, because it provides the smallest phase uncertainties for the time-delay analysis. Another option is to prewhiten the spurious frequencies from the time delays. This is more time-consuming and not easily automated, but results in a better determination of the orbital parameters and is the preferred option. That is not to say that the influence of spurious peaks can be entirely mitigated in practice. At some amplitude these cannot be easily distinguished from white noise, and so the general effect is an addition to the overall noise level.

\subsubsection{Detection limits and discussion}

Photon noise in the light curve causes white noise in the time delays, and therefore affects the minimum detectable companion mass. If we assume photon noise to be the dominant noise source in real light curves, then detection thresholds will lie at lower companion masses for brighter stars. However, the important quantity is the signal-to-noise ratio of the oscillations, so faint stars with large oscillation amplitudes may still have lower detection thresholds than brighter pulsators with weaker oscillations. The number and the frequencies of the oscillation modes also matter. The white noise decreases approximately as $\sqrt{N}$, where $N$ is the number of oscillations used in the weighted average, but adding more modes of lower amplitudes leads to diminished returns because they contribute less to the weighted average. The oscillations differ from star to star, so the detection thresholds for real data must be examined on an individual basis. None the less, we give some approximate limits based on simulations (see also \citealt{comptonetal2016}).

The detection limits are based on a simulation containing $\ell=0$ to 4 modes (shown in bottom panel of Fig.\,\ref{fig:mode_crowding}), with the addition of Gaussian-distributed white noise of $\sigma=0.13$\,mmag per point. This corresponds to the photometric precision achieved for a \textit{Kepler} target with $K_p = 13.0$\,mag. The signal-to-noise ratios of the ten strongest oscillation modes lie between 150 and 270 before prewhitening, or 260 to 430 after prewhitening their spectral windows from the light curve. Once this is done, the highest noise peaks in the Fourier transform of the time delays are at about 2--3\,s, from which we conclude that a binary system with $a_1 \sin i / c > 5$\,s would be detectable.

\begin{figure}
\begin{center}
\includegraphics[width=0.475\textwidth]{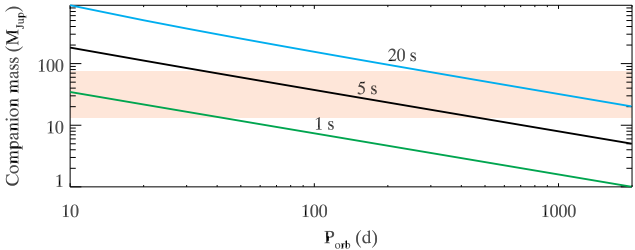}
\caption{Companion masses (in Jovian masses) corresponding to the detection limits in $a_1 \sin i / c$ for objects orbiting a typical non-radially pulsating $\delta$\,Sct star. The black line represents the detection limit of $a_1 \sin i / c = 5$\,s established from simulations; limits of 1\,s and 20\,s, for the best case and common case in \textit{Kepler} $\delta$\,Sct stars, are also shown. The canonical mass range of brown dwarfs is indicated by the shaded region. The primary is assumed to be 1.8\,M$_{\odot}$, and $\sin i = 1$.}
\label{fig:limits}
\end{center}
\end{figure}

When the noise is white, the detection limit (in terms of the minimum detectable value of $a_1 \sin i / c$) is the same at all periods. However, as seen in equation (\ref{eq:36}), the minimum mass corresponding to that $a_1 \sin i / c$ limit does depend on the orbital period because longer periods correspond to larger orbits (larger $a_1 \sin i / c$) for a companion of a given mass. Fig.\:\ref{fig:limits} provides the minimum detectable companion mass for a given $a_1 \sin i / c$ limit at different orbital periods.

From Fig.\,\ref{fig:limits} we conclude that the lowest-mass M dwarfs at 0.07\,M$_{\odot}$ (75\,M$_{\rm Jup}$) are detectable around $\delta$\,Sct stars even in unfavourable cases, if the orbital period exceeds 300\,d. M0 stars, with masses around 0.6\,M$_{\odot}$, are generally detectable at periods above 20\,d. These limits extend to lower mass (or shorter period) in more optimal cases. In the best conditions (high pulsation signal-to-noise), Jupiter-mass companions can be found at orbital periods comparable to the data set length. The sensitivity would have been sufficient to detect Jupiter and Saturn in their orbits, had \textit{Kepler} continued observing for a full orbit (i.e. 12\,yr for Jupiter). The mission actually lasted 1470\,d, which gives a practical limit of about 2--10\,M$_{\rm Jup}$, depending on the pulsation properties. Importantly, the periods at which planets are detectable includes the habitable zone.

Similar detection limits can be obtained for different classes of pulsating stars by scaling these results according to the oscillation frequencies and their signal-to-noise ratios (see \citealt{comptonetal2016}), and accounting for the difference in mass of the pulsators.

\subsection{Phase drift over time}

Small inaccuracies in the oscillation frequencies can cause a linear phase drift to accumulate over time \citep{sterken2005c}. This can be the case even if the oscillation frequencies are determined as accurately as possible within a given data set. If the frequency errors were random, the use of multiple pulsation modes would generally cancel out any such phase drift. However, we found that the oscillation frequencies could be systematically incorrect, and that this effect was strongly correlated with orbital period.

\begin{figure}
\begin{center}
\includegraphics[width=0.48\textwidth]{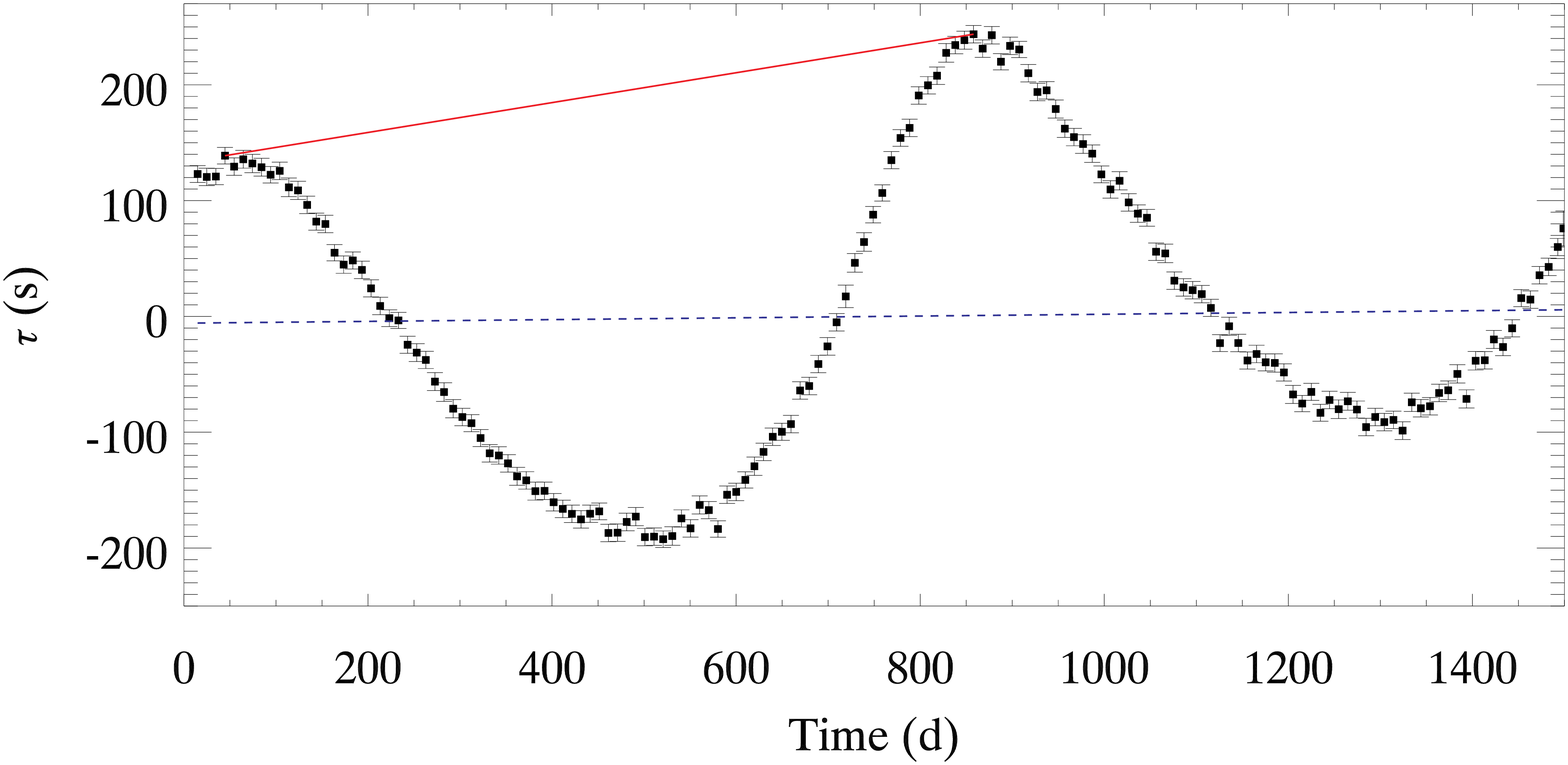}
\includegraphics[width=0.48\textwidth]{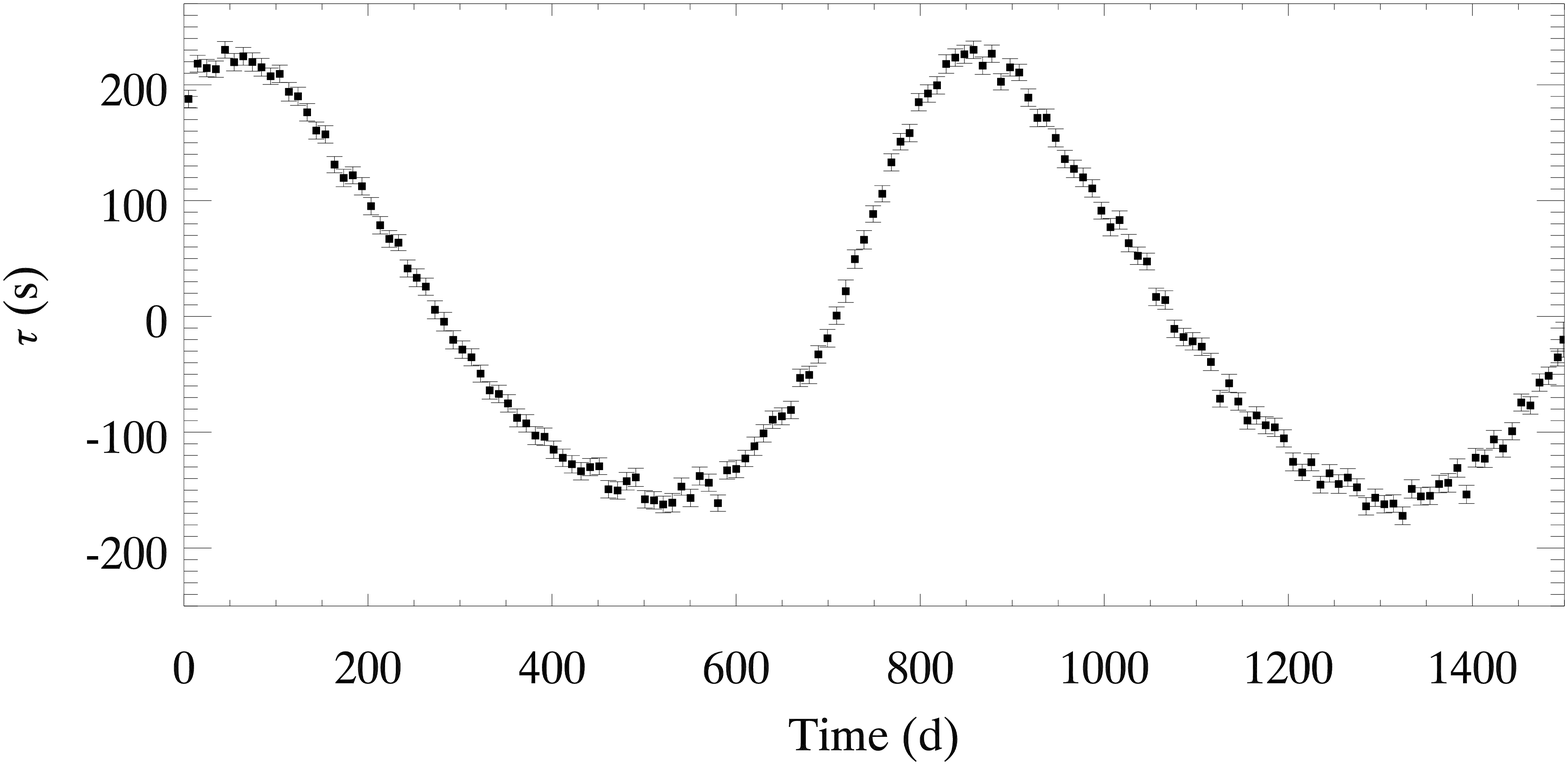}
\caption{Top: a linear drift in time delays, $\tau (t)$, caused by using a slightly incorrect oscillation frequency. The dashed blue line is a linear fit to all the data points, whereas the red line is a fit from peak-to-peak (see text). Bottom: the same time delays after a correction is applied.}
\label{fig:phase_drift}
\end{center}
\end{figure}

The cause is the time-averaged Doppler shift of the oscillation frequency, $\nu_{\rm osc}$, over the orbit. For an integer number of orbits, the observed oscillation frequency is equal to the input value. However, when the number of orbits is a small non-integer, the mean observed frequency is no longer equal to the input value, but is shifted by an amount
\begin{equation}
	\Delta \nu_{\rm osc} = \frac{(T~{\rm mod}~P_{\rm orb})}{T} \frac{\nu_{\rm osc}}{c} \int_0^T v_{\rm rad} (t)\, {\rm d}t,
\label{eq:30}
\end{equation}
where the coefficient $(T\,{\rm mod}\,P_{\rm orb})/T$ denotes the remainder of the division of the time span of the data $T$ by the orbital period $P_{\rm orb}$, and $v_{\rm rad} (t)$ is the instantaneous radial velocity of the star in its orbit. This produces a linear trend in the time delays as shown in Fig.\,\ref{fig:phase_drift}. This is almost certainly the cause of the linear trend in the O-C residuals for KIC\,11754974 found by \citet{murphyetal2013a}, rather than the difference in the \textit{Kepler} and WASP passbands suggested in that paper.

Attempts to phase-wrap time delays without correcting for a linear drift can result in large errors in the inferred orbital parameters. One cannot simply find and subtract a linear fit by least squares (dashed blue line, Fig.\,\ref{fig:phase_drift}) because the data do not cover an integer number of orbits. Instead, a correction can be applied by finding the gradient of a line connecting the maxima or minima in the time delays. The orbital parameters extracted after the correction is applied match the input values satisfactorily.

Phase drift of this kind is greater for longer orbital periods but the top-right panel of Fig.\,\ref{fig:oscillation_noise} shows it can be significant for periods of $\sim$300\,d. One cannot know, {\it a priori}, the value of the shift without an ephemeris for the orbit.

If two maxima or two minima are available then the time delays are easily detrended (Fig.\,\ref{fig:phase_drift}, bottom). If those maxima or minima have substantial scatter, then choosing the `correct' gradient can be arbitrary and can have significant consequences for the determined orbital parameters that is not reflected in the numerical uncertainties. For this reason, we included the gradient of the correction as an additional parameter in the MCMC analysis. We simulated several binary orbits with periods longer than 1000\,d and we were able to reproduce the input parameters more accurately when the detrending was incorporated within the MCMC framework than when it was applied manually.

\begin{table}
\centering
\caption{Comparison between the simulated and inferred parameters for a long-period binary system, where the orbital period twice as long as the 1500-d data set and a significant trend is seen in the time delays. The final column shows that the input values are reproduced at the 1--2$\sigma$ level (calculated as inferred minus input). Importantly the mass function is accurately reproduced.}
\label{tab:data_set_length}
\begin{tabular}{c c c c c}
\toprule
\multicolumn{1}{c}{Parameter} & units & \multicolumn{1}{c}{Input} & \multicolumn{1}{c}{Inferred} & Difference ($\sigma$)\\
\midrule
\vspace{1.5mm}
$P_{\rm orb}$ & d & 3017.1  & $3364^{+119}_{-322} $& +1.08 \\
\vspace{1.5mm}
$a_1 \sin i / c$ & s & 589.01 & $630^{+14}_{-37}$ & +1.07 \\
\vspace{1.5mm}
$e$ & & 0.75 & $0.768^{+0.010}_{-0.023}$ & +0.78 \\
\vspace{1.5mm}
$\phi_{\rm p}$ & $[$0--1$]$ & 0.1699 & $0.1539^{+0.0155}_{-0.0049}$ & $-1.03$ \\
\vspace{1.5mm}
$\varpi$ & rad & 1.6179 & $1.653^{+0.030}_{-0.015}$ & +2.34 \\
\vspace{1.5mm}
$f(m_1,m_2,\sin i)$ & M$_{\odot}$ & 0.0241 & $0.0238^{+0.0048}_{-0.0046}$ & $-0.06$ \\
\bottomrule
\end{tabular}
\end{table}

We also simulated binary orbits longer than the 1500-d data set length. We were able to recover the input parameters for periods up to around $P_{\rm orb} = 2000$\,d, with the agreement becoming poorer at longer periods. Degeneracies between the orbital period, the eccentricity and the gradient applied to the time delays generally prevent unique solutions from being found at longer periods, but some orbital configurations are still solvable. An example for a 3000-d binary is given in Table\:\ref{tab:data_set_length}. Importantly, since inaccuracies in $P_{\rm orb}$ and $a_1 \sin i / c$ are correlated (larger orbits have longer periods), the mass function is still accurately recovered. The consequence is that we may still identify low-mass companions such as brown dwarfs and planets when the orbital period is longer than the data set, where our sensitivity to such low-mass companions is highest.

\section{Incorporation of Radial Velocities}

\label{sec:rv}

Radial velocity curves for faint targets require many hours of time on large telescopes. In addition, it can be logistically difficult to obtain adequate orbital phase coverage for a sample of binary stars with a wide range of orbital periods. This makes the PM method, as applied to \textit{Kepler} data, a much more efficient survey for binary systems, where thousands of stars can be studied with little dependence on their visual magnitudes.

RV measurements can be imprecise for B/A/F-type stars due to their rapid rotation \citep{abt&morrell1995,royeretal2007}, while the PM method is well suited to pulsators at these spectral types \citep{comptonetal2016}. If the RV precision is poor, the small RV amplitudes of long period binaries may go undetected, whereas the PM method is more efficient at longer periods. On the other hand, the near-instantaneous nature of RV measurements (compared to time-delay integrations of several days) make RVs better suited to parametrizing compact orbits with small separations at periastron. Those include short-period systems as well as some with longer periods but high eccentricities, where the time delay changes rapidly at periastron.

\begin{figure}
\begin{center}
\includegraphics[width=0.49\textwidth]{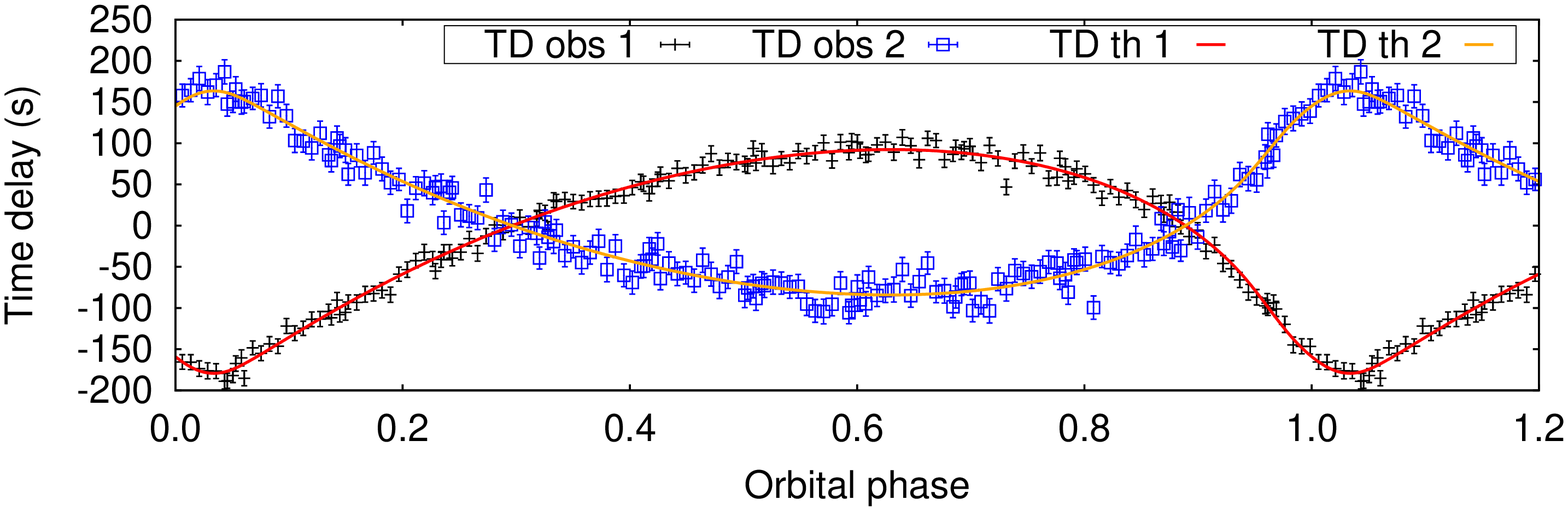}
\includegraphics[width=0.49\textwidth]{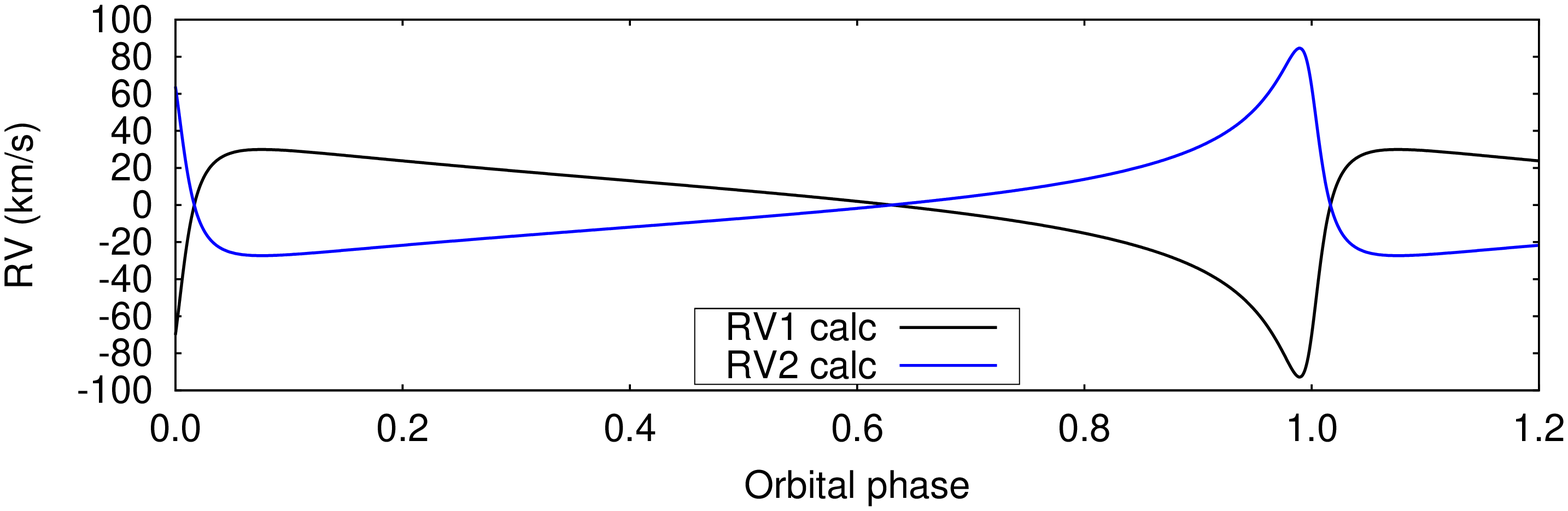}
\caption{A simulated PB2 system with input values of $P_{\rm orb}$ = 100\,d, $e=0.8$ and $m_1/m_2 = 0.9$. The RV curve for each component (bottom) can be computed from the best fitting theoretical orbit to the observed time delays (top) after correction for undersampling.}
\label{fig:pb2}
\end{center}
\end{figure}

Given that the PM method is relatively new, while RVs have been used for solving binary orbits for over a century, we produce RV curves for familiarity. We computed these from the orbital parameters determined by the MCMC method. An example PB2 time delay curve with a solved orbit is shown in Fig.\,\ref{fig:pb2}, along with the RV curve corresponding to the same orbit. Thus the semi-amplitude of the radial velocity, $K_1 := (v_{\rm rad, 1, max} - v_{\rm rad, 1, min})/2$, which is a key output of RV analyses, can be obtained analytically from PM orbits
\begin{equation}
	K_1 = \frac{(2\uppi G)^{1/3}}{\sqrt{1-e^2}} \left\{ {{ f(m_1,m_2,\sin i) }\over{P_{\rm orb}}} \right\}^{1/3},
\label{K1b}
\end{equation}
even when no RV observations are available.

\subsection{Inclusion of radial velocity data}

For many systems both the RV and PM methods can be useful. The criterion is that the measurement uncertainties are a small fraction of the total variation across the orbit.

We have developed a way to use both RVs and time delays in the MCMC framework for calculating the orbital parameters. The $\chi^2$ of a given orbit becomes the sum of the contributions from the fit to the time-delay data and the fit to the RV data. This requires real-time correction for undersampling by the time delays of each proposed orbit in the Markov chain (see Sect. \ref{ssec:undersampling}).

The two methods are highly complementary, and the result is a substantial refinement of the orbital parameters. RV observations made now (in the year 2016) will nearly double the baseline of observations for each binary system, given that the 4\,yr of \textit{Kepler} observations ended in 2013~May.

\subsection{Application to real data: PB1--SB1 systems}
\label{ssec:real_data}

\begin{figure}
\begin{center}
\includegraphics[width=0.495\textwidth]{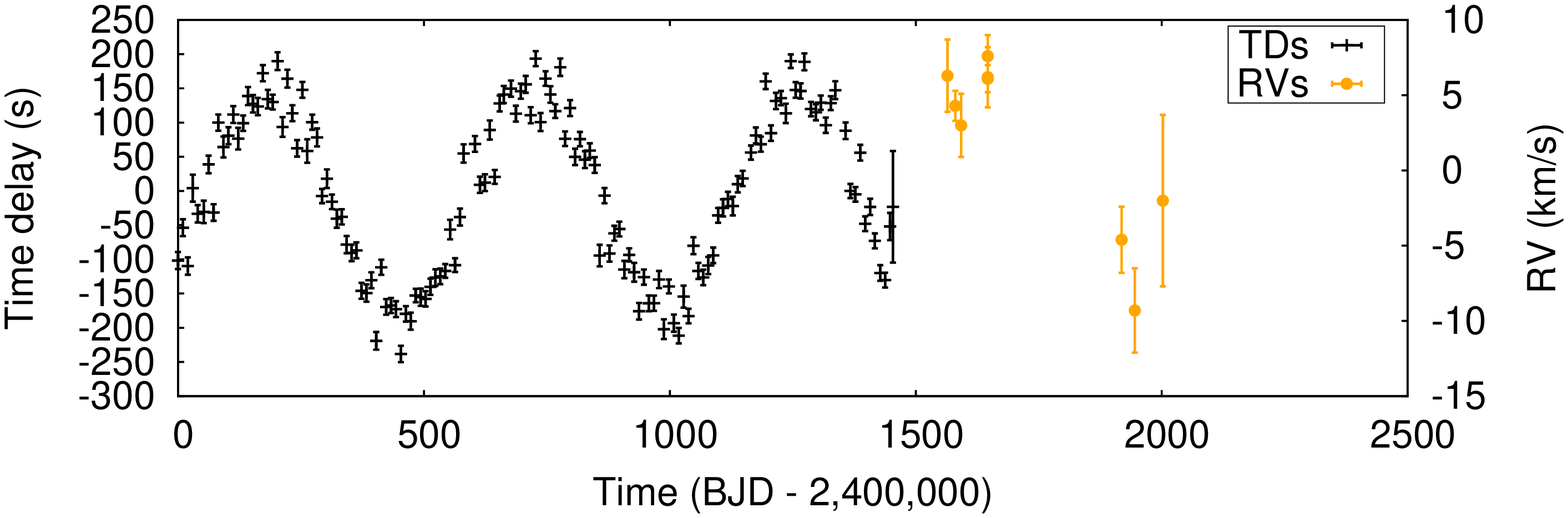}
\includegraphics[width=0.495\textwidth]{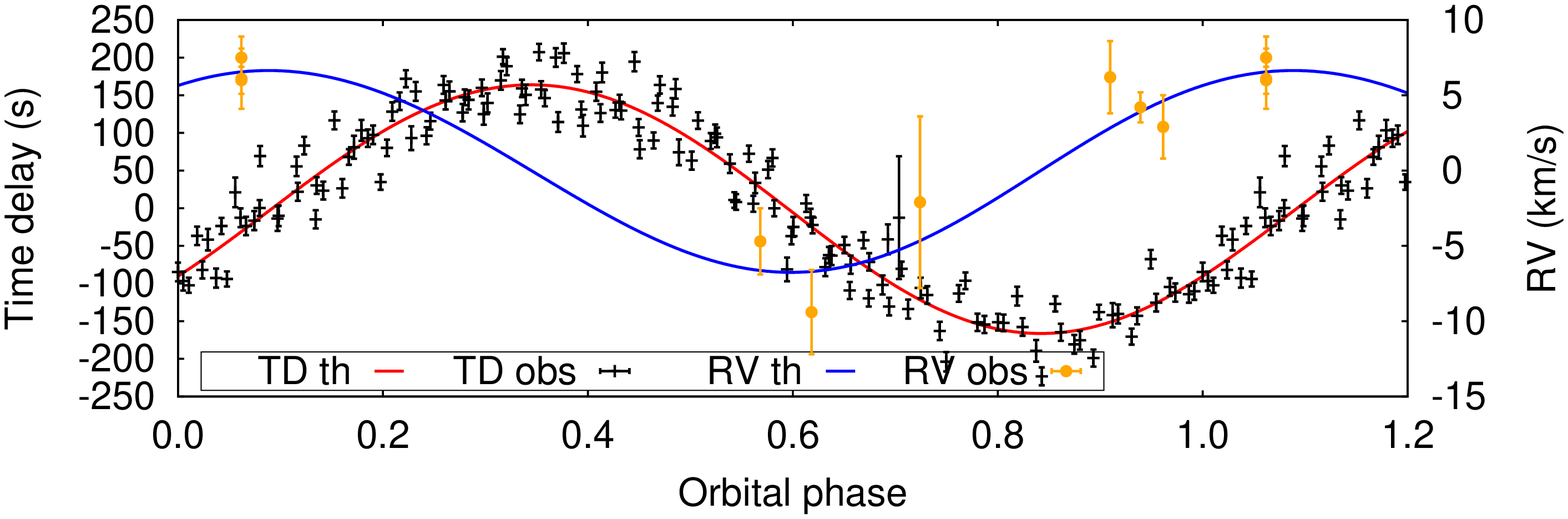}
\caption{Time delays (left vertical axis), and RVs (right vertical axis), as a function of time (top panel) and orbital phase (bottom panel) for the 537.7-d binary KIC\,5705575. Solid lines show the orbital solution. The seven pulsation modes used in the PM analysis are those with frequencies at 20.54, 26.67, 20.28, 25.28, 21.36, 24.78, and 25.83\,d$^{-1}$.}
\label{fig:integrated_kic5}
\end{center}
\end{figure}

\begin{figure}
\begin{center}
\includegraphics[width=0.495\textwidth]{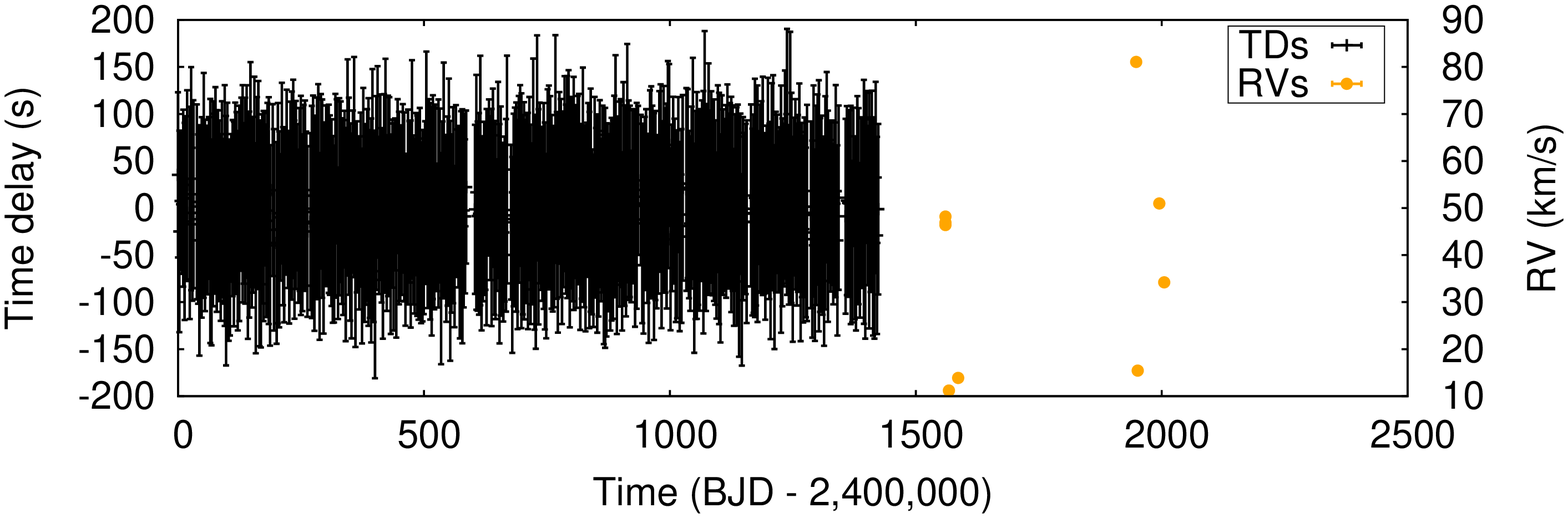}
\includegraphics[width=0.495\textwidth]{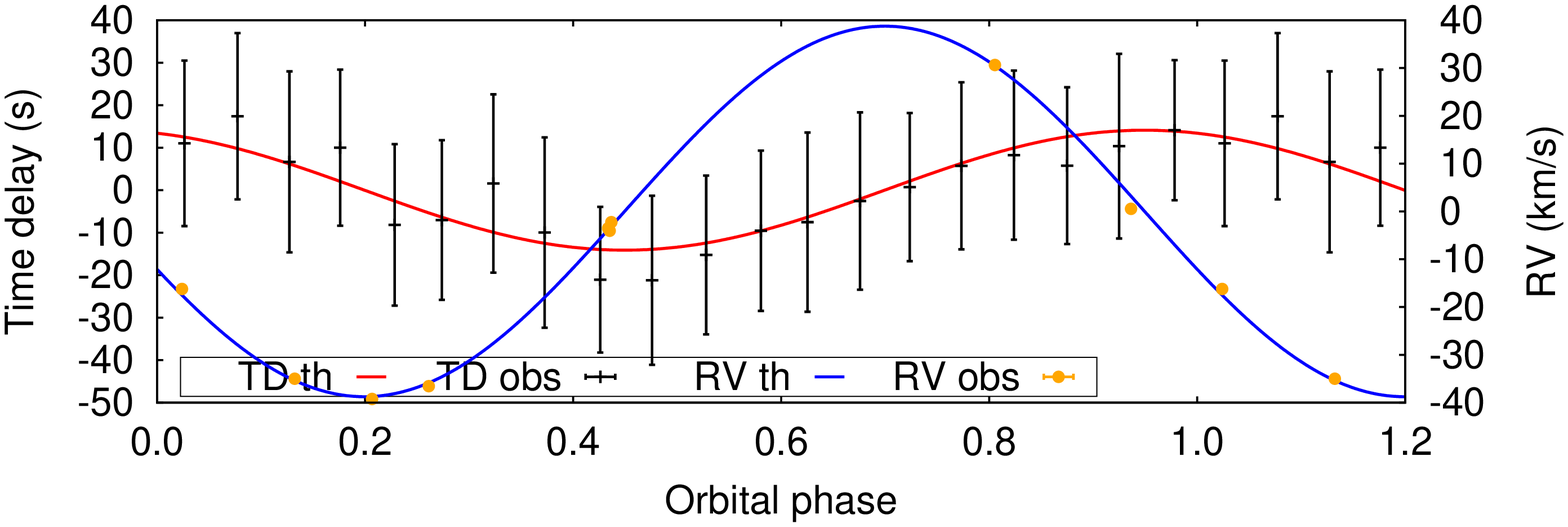}
\caption{Data coverage and phased orbital solution for KIC\,6780873, whose orbital period is 9.15\,d. Time delays from the only two strong modes, at 14.19 and 13.44\,d$^{-1}$, were used. Symbols and colours are the same as those in Fig.\,\ref{fig:integrated_kic5}.
}
\label{fig:integrated_kic6}
\end{center}
\end{figure}

The precision of the orbital parameters can sometimes be improved greatly with the inclusion of only a small number of RV measurements. Examples combining RVs and time delays are shown in Figs\,\ref{fig:integrated_kic5} and \ref{fig:integrated_kic6} for two \textit{Kepler} SX\,Phe binaries \citep{nemecetal2015}. For KIC\,5705575, the long orbital period of $537.7\pm0.9$\,d leads to large RV error bars relative to the RV semi-amplitude, so the few RV data do not add to the quality of the solution. For KIC\,6780873, which has a much shorter orbital period of just $9.1547\pm0.0003$\,d, the physical size of the orbit is small and the uncertainty on the time delays almost equals the peak-to-peak variation, but the time delay measurements are numerous. For this system, the combining time delays and RVs leads to substantial improvement in the orbital solution (Table\:\ref{tab:PMRV}). The eccentricity is improved by a factor of 50 compared to the solution from RVs alone. The period uncertainty of just 25\,s and the uncertainty in $a_1 \sin i/c$ of just 44\,ms is remarkable for a non-eclipsing and non-exotic binary (i.e., where neither star is a pulsar).

\begin{table}
\caption{Comparison of the precision on the orbital parameters for KIC\,5705575 and KIC\,6780873 when using different data sets. PM analyses used seven and two pulsation modes, respectively. The values of $\phi_{\rm p}$ and $\varpi$ are not shown because they are undefined for the small eccentricities. $P_{\rm orb}$ and $e$ values for the RV solution of KIC\,5705575 were fixed to the bracketed values because they could not be determined independently.}
\label{tab:PMRV}
\begin{tabular}{c c c r@{ $\pm$ }l c}
\toprule
\multicolumn{1}{c}{Parameter} & units & \multicolumn{1}{c}{PM only} & \multicolumn{2}{c}{RV only} & \multicolumn{1}{c}{PM + RV} \\
\midrule
\multicolumn{6}{c}{\it KIC\,5705575}\\
\midrule
\vspace{1.5mm}\hspace{-2mm}
$P_{\rm orb}$ & d & $537.5^{+0.9}_{-1.1}$  &  \multicolumn{2}{c}{(537.5)}  & $537.7^{+0.8}_{-0.9} $ \\
\vspace{1.5mm}\hspace{-2mm}
$a_1 \sin i / c$ & s & $165.4^{+1.4\phantom{0}}_{-1.3\phantom{0}}$ & 162 & 20 & $165.2^{+1.2}_{-1.3}$ \\
\vspace{1.5mm}\hspace{-2mm}
$e$ & & $\phantom{0}0.032^{+0.015}_{-0.016}$ &  \multicolumn{2}{c}{(0.0)}  & $0.017^{+0.015}_{-0.010}$ \\
\vspace{1.5mm}\hspace{-2mm}
$f(m_1,m_2,\sin i)$\hspace{-4mm} & M$_{\odot}$ & $0.0168^{+0.0004}_{-0.0004}$ & 0.016 & 0.006 & $0.0168^{+0.0003}_{-0.0004}$ \\
\vspace{1.5mm}\hspace{-2mm}
$K_1$ & km\,s$^{-1}$ & $6.71^{+0.04}_{-0.04}$ & 6.55 & 0.83 & $6.70^{+0.04}_{-0.03}$ \\
\midrule
\multicolumn{6}{c}{\it KIC\,6780873}\\
\midrule
\vspace{1.5mm}\hspace{-2mm}
$P_{\rm orb}$ & d & $9.131^{+0.025}_{-0.013}$  & 9.161 & 0.001 & $9.1547^{+0.0003}_{-0.0003} $ \\
\vspace{1.5mm}\hspace{-2mm}
$a_1 \sin i / c$ & s & $16.6^{+2.0\phantom{0}}_{-2.6\phantom{0}}$ & 16.5 & 0.5 & $16.278^{+0.042}_{-0.045}$ \\
\vspace{1.5mm}\hspace{-2mm}
$e$ & & $\phantom{0}0.034^{+0.024}_{-0.023}$ & 0.04 & 0.02 & $0.0004^{+0.0006}_{-0.0002}$ \\
\vspace{1.5mm}\hspace{-2mm}
$f(m_1,m_2,\sin i)$\hspace{-4mm} & M$_{\odot}$ & $0.059^{+0.021}_{-0.028}$ & 0.055 & 0.002 & $0.0553^{+0.0004}_{-0.0004}$ \\
\vspace{1.5mm}\hspace{-2mm}
$K_1$ & km\,s$^{-1}$ & $39.6^{+5.6}_{-5.6}$ & 38.7 & 0.9 & $38.77^{+0.06}_{-0.06}$ \\
\bottomrule
\end{tabular}
\end{table}

\subsection{Application to real data: a PB2--SB2 system}

\begin{figure}
\begin{center}
\includegraphics[width=0.5\textwidth]{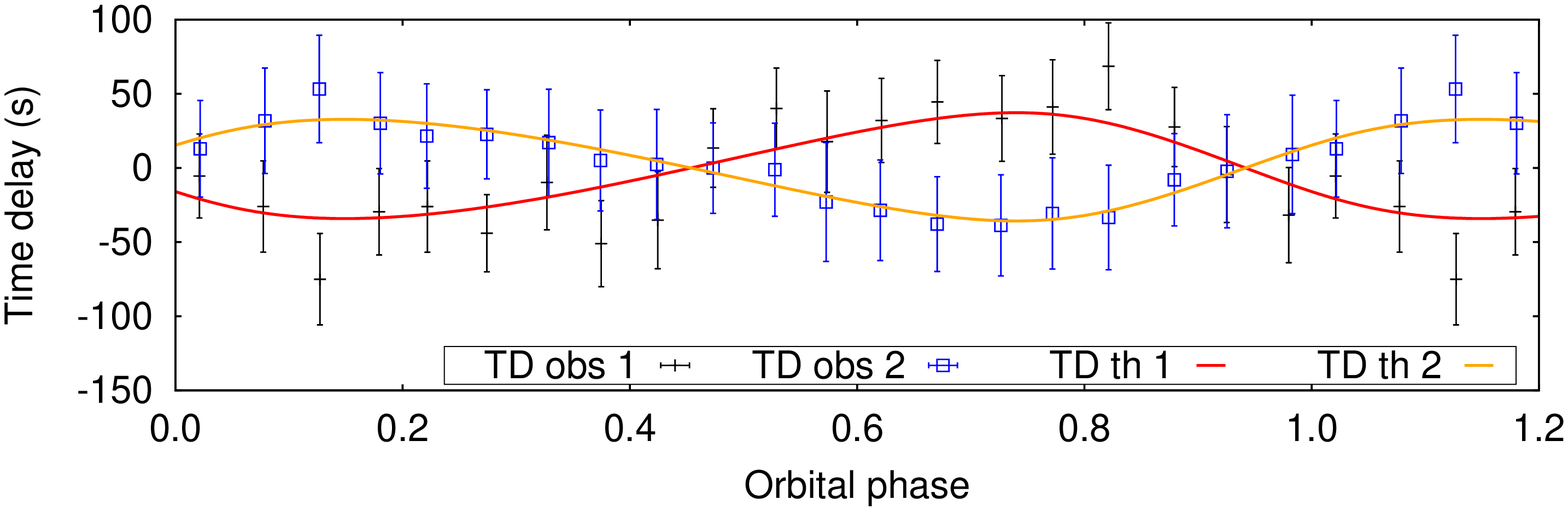}
\includegraphics[width=0.5\textwidth]{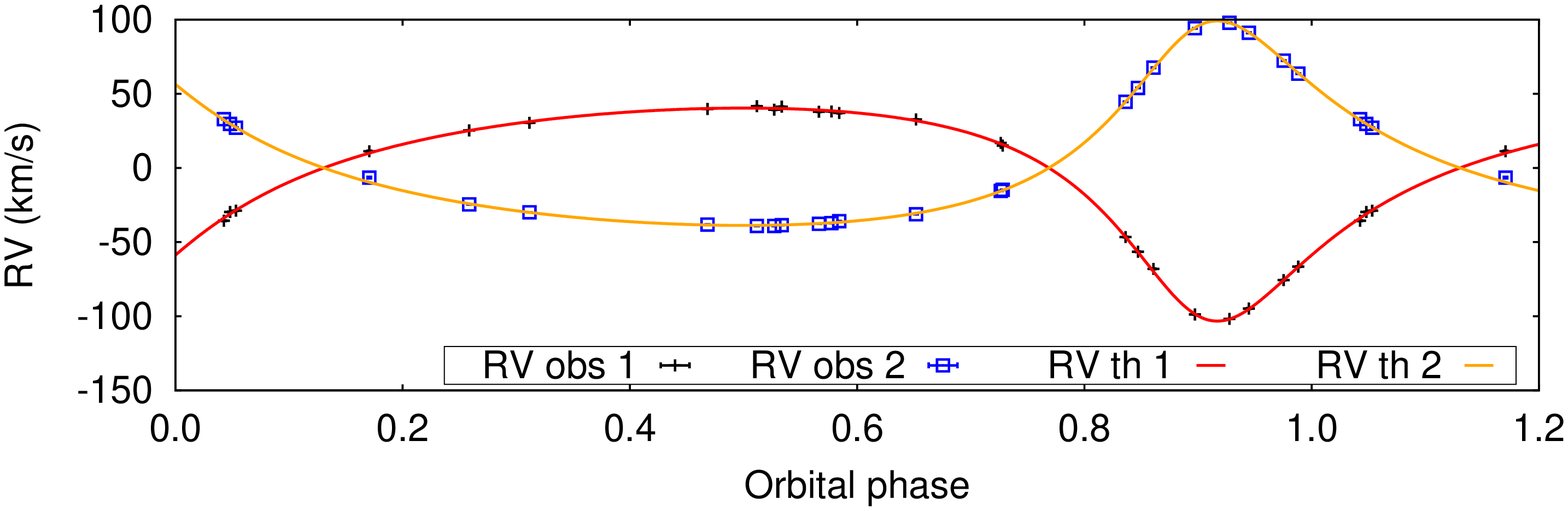}
\caption{Combined time delay (top) and RV (bottom) analysis of the double-lined system KIC\,10080943. The star of origin could be established for five pulsation modes with the PM analysis (see \citealt{schmidetal2015}), with frequencies of 13.95, 15.68, 12.89, 14.20 and 19.64\,d$^{-1}$. Theoretical orbits used for the RVs and time delays of a given star are the same. The time delay analysis used 80 bins but for clarity only 20 bins are shown.
}
\label{fig:SB2}
\end{center}
\end{figure}

The ability to derive the mass ratio of stars in a binary system independently of the inclination is a great advantage of studying double-lined systems, but it comes at the expense of disentangling the spectral or pulsational contributions from the two stars. When combining time delays and RVs, an additional complication can arise in associating the correct RV curve with the corresponding time-delay curve. Fig.\,\ref{fig:SB2} shows a PB2--SB2 system where time delays and RVs could be derived for both components. The system is KIC\,10080943, in which both stars are $\delta$\,Sct--$\gamma$\,Dor hybrids \citep{keenetal2015,schmidetal2015,schmid&aerts2016}. We used the RVs from \citet{schmidetal2015} in our analysis here.

The short orbital period of 15.33\,d is much more favourable to RVs than to time delays for deriving the orbital parameters. Additionally, the number of RV measurements and their phase coverage produce a good solution on their own, as \citet{schmidetal2015} showed with their analysis using the FDBinary code \citep{ilijicetal2004}. None the less, the time-delay data do improve the solution, as shown in Table\:\ref{tab:SB2}. The time delays provide a longer observational timespan, which improves the precision on the orbital parameters by a factor of 10. For this double-lined system, we have determined the orbital period to a precision of 3\,s.

\begin{table}
\centering
\caption{Improvement in the orbital parameters for KIC\,10080943, based on the combination of time-delay and RV data. Five oscillation modes could be used for the PM analysis. Column 3 refers to the parameters published by \citet{schmidetal2015}. The value of $t_{\rm p}$ (in units of BJD$-2\,400\,000$) is given instead of $\phi_{\rm p}$, but was forward-calculated to the epoch of the \citet{schmidetal2015} value, resulting in uncertainties that are larger than the smallest achievable uncertainties of 0.0003\,d for this system.}
\label{tab:SB2}
\begin{tabular}{c c r@{ $\pm$ }l c}
\toprule
\multicolumn{1}{c}{Parameter} & units & \multicolumn{2}{c}{\hspace{4mm}RV only} & \multicolumn{1}{c}{PM + RV} \\
\midrule
\vspace{1.5mm}\hspace{-2mm}
$P_{\rm orb}$ & d & 15.3364 & 0.0003 & $15.33619^{+0.00004}_{-0.00004} $ \\
\vspace{1.5mm}\hspace{-2mm}
$t_{\rm p}$ & d & 55\,782.23 & 0.02 & $55\,782.242^{+0.018}_{-0.018\phantom{0000}}$\\
\vspace{1.5mm}\hspace{-2mm}
$a_1 \sin i / c$ & s & 43.0 & 0.3 & $43.22^{+0.02}_{-0.02}$ \\
\vspace{1.5mm}\hspace{-2mm}
$a_2 \sin i / c$ & s & 44.7 & 0.3 & $45.02^{+0.02}_{-0.02}$ \\
\vspace{1.5mm}\hspace{-2mm}
$e$ & & 0.449 & 0.005 & $0.4539^{+0.0003}_{-0.0003}$ \\
\vspace{1.5mm}\hspace{-2mm}
$\varpi$ & rad & 6.016 & 0.012 & $6.0187^{+0.0010}_{-0.0008}$ \\
\vspace{1.5mm}\hspace{-2mm}
$f(m_1,m_2,\sin i)$ & M$_{\odot}$ & 0.3628 & 0.0076 & $0.3687^{+0.0006}_{-0.0006}$ \\
$q$ & & 0.96 & 0.01 & $0.960\pm0.001$ \\
\bottomrule
\end{tabular}
\end{table}



\section{Conclusions}

We have developed MCMC software to solve binary orbits based on a series of time delay observations, and to provide robust uncertainties. We simulated orbits covering a range of parameters to explore the sensitivity limits of the method, the factors governing those limits, and to predict the lowest mass companions detectable by the method. We confirmed that the method is much more sensitive to stars oscillating with high signal-to-noise, and in such cases the detection limit approaches 1--2\,M$_{\rm Jup}$ at long orbital periods ($>1000$\,d), where the habitable zones of intermediate-mass stars are located.

We also showed that orbital solutions can be obtained when the orbital period is longer than the data timespan, with the upper limit on orbital period depending on the orientation and eccentricity of the orbit, and whether or not the periastron phase is observed. The uncertainties on such orbits tend to be large and perhaps underestimated. However, since any overestimates of the orbital period will be correlated with overestimates of $a_1 \sin i / c$, and vice-versa, the mass function is well recovered. This is because the rate of change of the time delays, which is governed by the mass function, can be established without observing a full orbit.

One drawback to the PM method is that we must divide the light curve into segments of several days to make adequate measurements of the pulsation phases. For short-period binaries this leads to significant undersampling and if the orbits are eccentric, the time-delay curve is heavily smeared. We have overcome this drawback by developing correction factors to the time-delay fitting function, and we verified the validity and implementation of that function in the MCMC algorithm via a hare and hounds exercise. The undersampling correction can help in understanding our completeness in surveys for binary stars with short periods, but a full completeness analysis remains as future work.

Another development is the simultaneous use of radial velocities and time delays as input data for solving the orbits in the MCMC framework. This requires implementation of an undersampling correction the time-delay fitting function, but allows the orbital parameters to be determined much more precisely. This is partly because of the complementarity of the PM and RV methods: the latter is the time derivative of the former and measurement of both provides a clear improvement in constraints on the orbit. Additionally, any RV measurements made now will double the time span of the observations, constraining the orbital period much more tightly, and thereby reducing the uncertainties on the other orbital parameters. For a real \textit{Kepler} binary system we showed that the combination of RVs and time delays can constrain the eccentricity to a factor 50 better than either the time delays or RVs alone, and orbital periods can be measured with a precision of seconds, even without eclipses.

\bibliography{sjm_bibliography} 

\section*{Acknowledgements}

We are grateful to J.~Nemec for providing his RV measurements of some \textit{Kepler} binaries ahead of publication, and to V.~Schmid for supplying the RVs used in her paper on KIC\,10080943. We thank T.~Sonoi for computing the stellar models upon our request, and J.~Hamann for discussions on the implementation of the Metropolis-Hastings algorithm. This research was supported by the Australian Research Council. Funding for the Stellar Astrophysics Centre was provided by the Danish National Research Foundation (grant agreement no.: DNRF106). The research was supported by the ASTERISK project (ASTERoseismic Investigations with SONG and Kepler) funded by the European Research Council (grant agreement no.: 267864). The research was also partially supported by the Japan Society for the Promotion of Science Grants-in-Aid for Scientific Research (grant no.: 16K05288).
SJM is an International Research Fellow of the Japan Society for the Promotion of Science.

\end{document}